\documentclass[iop]{emulateapj}

\def\gs{\mathrel{\raise0.35ex\hbox{$\scriptstyle >$}\kern-0.6em
\lower0.40ex\hbox{{$\scriptstyle \sim$}}}}
\def\ls{\mathrel{\raise0.35ex\hbox{$\scriptstyle <$}\kern-0.6em
\lower0.40ex\hbox{{$\scriptstyle \sim$}}}}

\makeatother

\righthead{An ALMA survey of U/LIRGs in XCS\,J2215 at \emph{z}\,=\,1.46}
\lefthead{Stach et al.}

\begin{document}

\title{ALMA pin-points a strong over-density of U/LIRGs in the massive cluster XCS\,J2215 at \emph{z}\,=\,1.46}


\author{Stuart M.\ Stach,$\!$\altaffilmark{1} A.\,M.\ Swinbank,$\!$\altaffilmark{1} Ian Smail,$\!$\altaffilmark{1} Matt Hilton,$\!$\altaffilmark{2}, J.\,M.\ Simpson$\!$\altaffilmark{3} E.\,A.\ Cooke$\!$\altaffilmark{1}}
\altaffiltext{1}{Centre for Extragalactic Astronomy, Department of Physics, Durham University, South Road, Durham, DH1 3LE, UK; email: stuart.m.stach@durham.ac.uk}
\altaffiltext{2}{Astrophysics and Cosmology Research Unit, School of Mathematics, Statistics and Computer Science, University of KwaZulu-Natal, Durban 4041, South Africa}
\altaffiltext{3}{Academia Sinica Institute of Astronomy and Astrophysics, No. 1, Sec. 4, Roosevelt Rd., Taipei 10617, Taiwan}

\setcounter{footnote}{1}

\begin{abstract}
We have surveyed the core regions of the $z=$\,1.46 cluster
XCS\,J2215.9$-$1738 with the Atacama Large Millimeter Array (ALMA) and
the MUSE-GALACSI spectrograph on the VLT.  We obtained high spatial
resolution observations with ALMA of the 1.2\,mm dust continuum and
molecular gas emission in the central regions of the cluster. These
observations detect 14 significant millimetre sources in a region with
a projected diameter of just $\sim$\,500\,kpc ($\sim$\,1$'$).  For six
of these galaxies we also obtain $^{12}$\,CO(2--1) and
$^{12}$\,CO(5--4) line detections, confirming them as cluster members,
and a further five of our millimetre galaxies have archival
$^{12}$CO(2-1) detections which also place them in the cluster.  An
additional two millimetre galaxies have photometric redshifts
consistent with cluster membership, although neither show strong line
emission in the MUSE spectra.  This suggests that the bulk
($\geq$\,11/14, $\sim$\,80\%) of the submillimetre sources in the
field are in fact luminous infrared galaxies lying within this young
cluster. We then use our sensitive new observations to constrain the
dust-obscured star formation activity and cold molecular gas within
this cluster. We find hints that the cooler dust and gas components
within these galaxies may have been influenced by their environment
reducing the gas reservoir available for their subsequent star
formation. We also find that these actively star-forming galaxies have
the dynamical masses and stellar population ages expected for the
progenitors of massive, early-type galaxies in local clusters
potentially linking these populations.
\end{abstract}


\section{Introduction} \label{sec:intro}

Galaxy clusters present a convenient laboratory for the study of
environmental influences on galaxy formation and evolution due to the
large variety in environments within a relatively small observable
area, from the high-density cores to the low-density outskirts.
Observational studies of clusters in the local universe show that
their cores are dominated by metal rich, gas-poor early-type
(lenticulars, or S0s, and ellipticals) galaxies with little or no
current star-formation activity. In contrast, late-type, actively
star-forming disk galaxies are found preferentially in the outskirts
of clusters and in the surrounding lower-density field, yielding a
so-called ``morphology--density'' relation
\citep{dressler1980galaxy,bower1992precision,whitmore1993determines,bamford2009galaxy}.

This correlation of galaxy star-formation activity and morphology with
environment in the local universe
\citep[e.g.][]{lewis20022df,gomez2003galaxy,balogh2004galaxy,kodama2004down}
is suggestive of environmental processes being at least partly
responsible for the quenching of star formation in the early-type
galaxies in high-density regions. Potential environmental processes
which could drive this include galaxies interacting with the
intracluster medium (ICM) causing ram pressure stripping of their
interstellar gas \citep{gunn1972infall}, or ``strangulation,'' where
the continued accretion of gas from their surroundings is cut off
\citep{larson1980evolution}; galaxy mergers leading to dramatic
changes in galaxy's structure and the triggering of a starburst which
rapidly consumes their gas \citep{merritt1983relaxation}; and tidal
interactions, which can enhance star formation
\citep{aguilar1985tidal}. Ultimately, each of these processes acts to
reduce the gas supply and eventually shut off star formation, and all
act preferentially on galaxies in higher density regions.

At higher redshift, it has been shown that the fraction of blue
star-forming disk galaxies found in clusters increases
\citep{butcher1978evolution,aragon1993evidence}.  A similar behaviour
has also been seen in these clusters using star-formation tracers
that are less sensitive to dust extinction, such as mid-infrared
emission. Indeed, 24\,$\mu$m surveys of actively star-forming galaxies
using the MIPS instrument on the {\it Spitzer Space Telescope} have
found increasing numbers of starbursts in clusters out to
$z\sim$\,0.5--1, although these clusters still typically contained a
core of passive galaxies
\citep[e.g.][]{geach2006panoramic,fadda2007starburst,saintonge2008spitzer,haines2009locuss,finn2010dust,biviano2011spitzer}.
The mass-normalized integrated star-formation rate (SFR) for these systems
increases with redshift as $\propto (1+z)^{\gamma}$ with $\gamma \sim
$\,7 \citep{geach2006panoramic}, an accelerated evolution in
comparison to the field where $\gamma \sim $\,4
\citep{cowie2004evolution}.

Although these clusters are growing at $z\sim$\,0.5--1, through
the infall and accretion of star-forming galaxies from their
environment, the cores still contain a population of massive, passive
galaxies which suggests that at least some of their galaxies must have
formed their stars at much earlier epochs. At higher redshifts,
  it has been shown that cluster cores show SFR
  densities equivalent to that of the field
  \citep{brodwin2013era,darvish2016effects} and in some $z\gtrsim1$
  cluster cores there are even claims of a reversal of the
  star-formation--density relation \citep[e.g.][]{tran2010reversal}.

However, one issue with these studies is that for clusters at $z>$\,1
the observed 24\,$\mu$m emission, which is often used as a
star-formation tracer, becomes increasingly problematic due to the
presence of strong, redshifted emission from polycyclic aromatic
hydrocarbon (PAH) and silicate absorption features which fall in the
band. As a result, studies of more distant clusters have focused on
the far-infrared/submillimetre wavebands and have uncovered evidence
of a continued rise with redshift in the activity in overdense
regions at $z>1$, as traced by an increasing population of the most
strongly star-forming, dusty (Ultra-)Luminous InfraRed Galaxies
(U/LIRGs)
\citep[e.g.][]{webb2005submillimeter,webb2013evolution,tran2010reversal,popesso2012evolution,smail2014scuba,noble2016phase}.
These studies have uncovered mixed evidence of a reversal in the
star-formation--density relation in cluster cores at high
redshift. For example, a ``reversal'' has been claimed in some massive
clusters at $z\gs$\,1.5 such as XDCP\,J0044.0$-$2033
\citep{santos2015reversal} and Cl\,J1001+0220
\citep{wang2016discovery}, but this is not
ubiquitous. \cite{smail2014scuba} identify 31 probable cluster U/LIRGs
within Cl\,J0218.3$-$0510 at $z=$\,1.62. However, these highly
  star-forming galaxies did not reside in the densest regions of the
  cluster and instead the core was already populated with passive red
  galaxies, a trend also seen by \cite{newman2014spectroscopic} in a
  $z=$\,1.8 cluster, which has a cluster core dominated by a quiescent
  galaxy population. These results suggest that a massive quiescent
  population in some $z\sim$\,1.5 cluster cores is already in place
  well before this epoch.

Some of the disagreement between the conclusions of these various
studies may result from the uncertainties in reliably identifying the
counterparts of far-infrared/submillimetre sources at other
wavelengths, due to the typically poor spatial resolution of the
long-wavelength data from single-dish facilities. To make progress on
these issues, we took high-spatial resolution millimetre imaging
of one of the well-studied high-redshift cluster which appears to
exhibit a very significant overdensity of submillimetre sources in
its core: XCS\,J2215.9$-$1738 \citep{stanford2006xmm}.  This cluster
has been claimed to exhibit enhanced star-formation activity in its
core regions \citep{hayashi2010high}, including a striking
overdensity of submillimetre sources in single-dish 450/850\,$\mu$m
maps obtained by \cite{ma2015dusty}.

In this paper, we present Atacama Large Millimeter/submillimeter Array
(ALMA) interferometric observations of dust continuum and CO emission
of galaxies in the central region of XCS\,J2215.  Our observations
include a 1.2\,mm mosaic of a 500\,kpc diameter region encompassing
the central four SCUBA-2 850\,$\mu$m sources detected by
\cite{ma2015dusty} (hereafter, Ma15). Our ALMA data provide us with
the means to study the U/LIRG population in this cluster in the
millimetre at resolutions an order of magnitude higher than that
provided from current single-dish bolometer cameras and with much
greater sensitivity. We use our ALMA continuum observations to
robustly identify the 850\,$\mu$m counterparts. We then searched for
emission lines arising from molecular gas in cluster members.  At the
cluster redshift, our ALMA observations in Band 3 and Band 6 cover two
transitions commonly seen in star-forming galaxies: $^{12}$CO(2--1)
and $^{12}$CO(5--4). We employ these detections to confirm the cluster
membership of U/LIRGs seen toward the cluster core and to estimate
their molecular gas content and physical properties.

This paper is structured as follows: \S 2 covers the target selection,
the ALMA observations and data reduction, with the resultant continuum
and CO line detections reported in \S 3. We then discuss these in \S 4
and give our main conclusions in \S 5.  We assume a $\Lambda$CDM
cosmology with $\Omega_{M} =$\,0.3, $\Omega_{\Lambda} =$\,0.7 and $H_0
= $\,70\,km\,s$^{-1}$\,Mpc$^{-1}$, which gives an angular scale of
8.5\,kpc\,arcsec$^{-1}$ at $z = $\,1.46. We adopt a Chabrier initial
mass function (IMF) \citep{chabrier2003galactic} and any magnitudes
are quoted in the AB system.

\section{Observations and Data Reduction} \label{sec:data}

%
%
\begin{figure*}[tbh]
	\centerline{\includegraphics[width=5.5in]{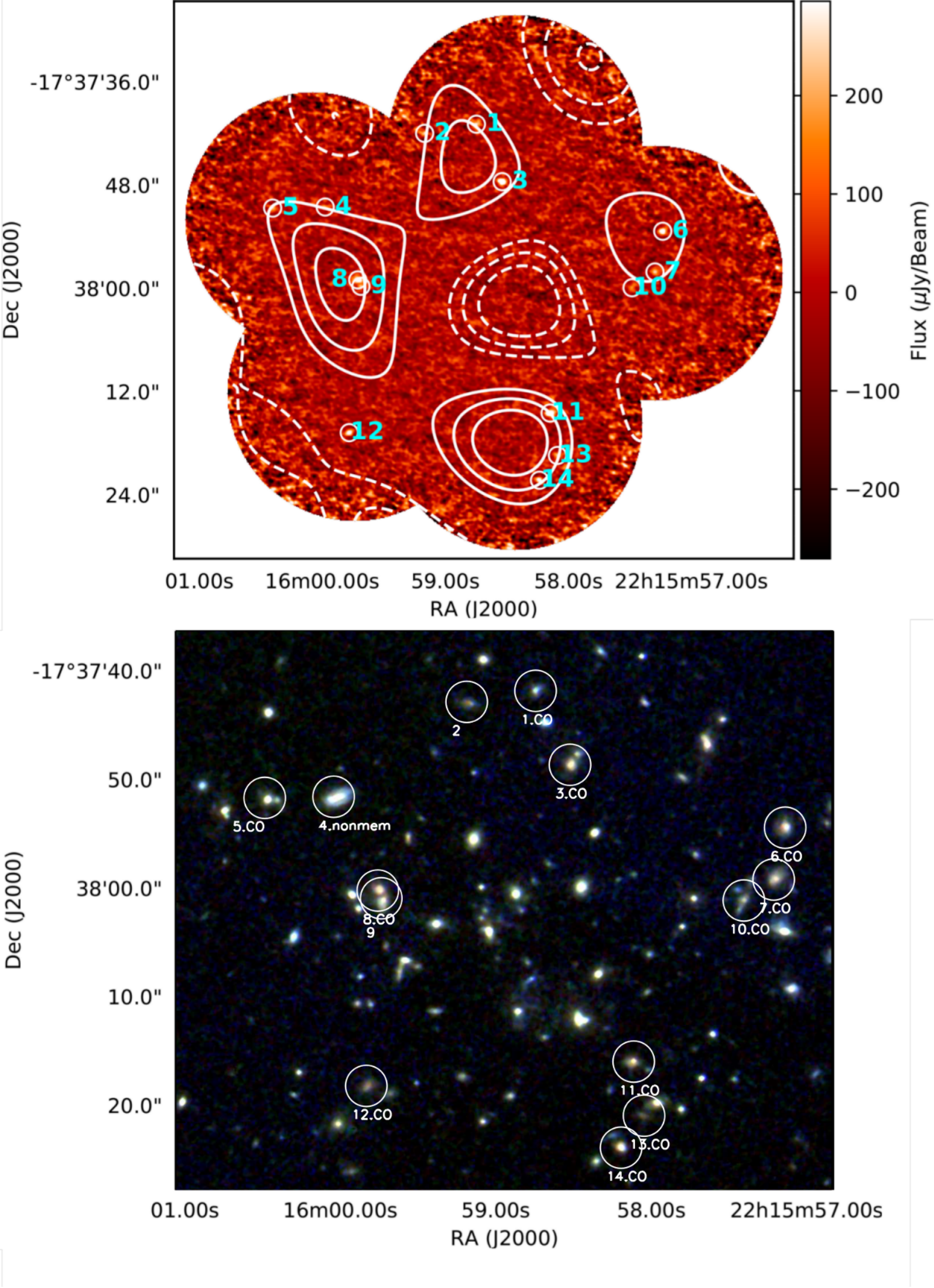}}
\caption{{\it Upper panel}: the ALMA 1.25\,mm (Band 6) mosaic of
  XCS\,J2215 taken from six overlapping pointings covering a 500\,kpc
  diameter region in the cluster core.  We detect 14 $>$4\,$\sigma$
  continuum detections, demonstrating a very significant overdensity
  of millimetre sources in this region marked by circles and
  numbered. We list their properties in Table~\ref{tab:results}. We
  also overlay the SCUBA-2 850\,$\mu$m S/N contours from Ma15 starting
  at 2\,$\sigma$ and increasing in steps of 1\,$\sigma$ (dashed lines
  showing the equivalent negative contours). {\it Lower panel}: a
  slightly zoomed three-colour {\it HST} image (F125W, F140W, and
  F160W), with our ALMA detections labelled, showing the rest-frame
  $V$-band morphologies of the millimetre sources. We highlight source
  \#4 as ``nonmem'' as this is a known interloper from its
  spectroscopic redshift; all the other sources have either $^{12}$CO
  (labelled ``CO'') detections from this study or archival
  $^{12}$CO(2-1) which confirm cluster membership (eleven sources) or
  photometric redshifts suggesting possible membership (IDs 2, 9)}
    \label{fig:alma_fov}
\end{figure*}

\subsection{XCS\,J2215.9-1738}

XCS\,J2215 provides an excellent opportunity to study the nature of
star-formation activity in the central regions of a high-redshift
cluster. At $z=$\,1.46 it is one of the most distant clusters
discovered in X-rays \citep[]{stanford2006xmm}, with extensive
multiwavelength follow-up
\citep{hilton2007xmm,hilton2009xmm,hilton2010xmm,hayashi2010high,hayashi2014mapping}.
Of particular relevance here is the SCUBA-2 survey of the clusters by
Ma15 which discovered an overdensity of submillimetre galaxies (SMGs)
in its core. Unlike other (proto-)clusters studied at high redshifts
\citep[e.g. CLG\,J0218][]{rudnick2012tale,lotz2013caught,hatch2016impact},
XCS\,J2215 appears structurally well-developed. By combining
\textit{XMM-Newton} and \textit{Chandra} observations,
\cite{hilton2010xmm} (hereafter H10) derived an X-ray luminosity of
$L_{X} = $\,2.9$^{+0.2}_{-0.4}\times$\,10$^{44}$\,erg\,s$^{-1}$ and an
ICM temperature $T=$\,4.1$^{+0.6}_{-0.9}$\,keV. Employing the
  $R_{200}$--velocity dispersion relation of
  \cite{carlberg1997average}; where $R_{200}$ is the radius from the
  cluster centre within which the mean density is 200 times the
  critical density at the redshift of the cluster, H10 used an
  iterative method to estimate a line-of-sight velocity dispersion of
  $\sigma_v = $\,720$\pm$110\,km\,s$^{-1}$ from the 31 galaxies with
  spectroscopic redshifts within $R_{200} = $\,0.8$\pm$0.1\,Mpc or
  100\arcsec. The velocity distribution of the galaxies, however, did
  show signs of bimodality, suggesting that the cluster may not be a
  completely relaxed and virialized system.

Within the central 0.25\,Mpc of the cluster
\cite{hayashi2010high,hayashi2014mapping} found 20 [O{\sc ii}]
emitters with dust-free star-formation rates (SFR)
$>$\,2.6\,$M_{\sun}$\,yr$^{-1}$. Using \textit{Spitzer}/MIPS, H10 found
a further three bright 24-$\mu$m sources with estimated SFRs of
$\sim$\,100\,$M_{\sun}$\,yr$^{-1}$ within the central
0.25\,Mpc. However, as noted earlier, at $z=$\,1.46 the broad
PAH feature at 8.6\,$\mu$m and
potential silicate absorption features are redshifted into the
24\,$\mu$m MIPS band, complicating the measurements of SFRs
 from this mid-infrared band. To provide a more robust census of
luminous dusty galaxies Ma15 obtained sensitive, longer wavelength
observations with SCUBA-2 at 850/450\,$\mu$m of XCS\,J2215.  These
observations were combined with JVLA observations at 1.4\,GHz and
archival images and photometry from \textit{Hubble Space Telescope}
({\it HST}), Subaru, and \textit{Spitzer}
\citep[respectively:][]{dawson2009intensive,hilton2009xmm,hilton2010xmm}
to study the U/LIRGs in the cluster. Ma15 detected seven submillimetre
sources above a 4-$\sigma$ significance cut within $R_{200}$
(100\arcsec \,radius), an order of magnitude above the expected
blank-field counts. A further four fainter ($>$\,3$\sigma$)
850\,$\mu$m sources were detected which were confirmed through
counterparts in \textit{Herschel}/PACS 70$\mu$m, 160$\mu$m and MIPS
24$\mu$m.  The probabilistic identification of counterparts to these
submillimetre sources in the mid-IR and radio associated 9 of the 11 
with galaxies that had spectroscopic or photometric redshifts that
suggested that they are cluster members. The total SFR from these
potential U/LIRG cluster members yields an integrated SFR within
$R_{200}$ of $>$\,1400\,$M_{\sun}$\,yr$^{-1}$, this suggests that
XCS\,J2215 is one of the highest SFR clusters known at high redshifts
\citep{ma2015dusty}.

We note that after submission of this paper
  \cite{hayashi2017evolutionary} published an ALMA Band 3
  $^{12}$CO(2-1) study of XCS\,J2215 which overlaps with our
  observations. Their concentration on just $^{12}$CO(2-1) enabled
  them to take deeper integrations over a slightly wider field and
  hence allowed them to detect fainter sources; however, where our
  observations overlap we obtain similar results.

\subsection{ALMA Band 6 Observations}

We obtained 1.25\,mm continuum and simultaneous
$^{12}$CO($J$\,=\,5--4) observations of the core of the XCS\,J2215
cluster using ALMA covering the core of the cluster including four of
the SCUBA-2 sources identified by Ma15. These Band 6 observations were
carried out on 2016 June 19 (project ID: 2015.1.00575.S). To cover the
$^{12}$CO(5--4) emission lines we set two spectral windows (SPWs) to
cover the observed frequencies from 232.7 to 236.4\,GHz or
$\Delta$V\,$\sim$\,4800\,km\,s$^{-1}$ which comfortably covers the
expected 720$\pm$110\,km\,s$^{-1}$ velocity dispersion of the
cluster. A further two SPWs were centred at 248.9 and 251.4\,GHz,
where no visible emission lines are expected, for continuum
imaging. Each SPW had a bandwidth of 1.875\,GHz with a spectral
resolution of 3.904\,MHz for the emission-line SPWs (corresponding to
a velocity resolution of 4.97--5.01\,km\,s$^{-1}$) and a spectral
resolution of 31.250\,MHz for the continuum SPWs
(37.3--37.6\,km\,s$^{-1}$). At these frequencies the
full-width-half-maximum (FWHM) of the primary beam is
$\sim$\,25\arcsec; therefore, a mosaic of six pointings was required to
map the central 500\,kpc diameter covering the cluster core
(Fig. \ref{fig:alma_fov}). The observations were conducted with forty-two
12\,m antennae where the bandpass calibration was obtained from
J2258$-$2758, the flux calibrator used was Titan, and the phase
calibrator was J2206$-$1835 with an on-source integration time of
244\,s for each pointing.

Calibration and imaging was carried out with the \textsc{Common
  Astronomy Software Application} (\textsc{casa} v4.6.0
\citep{mcmullin2007casa}). The observation used a configuration which
yielded a synthesised beam in Band 6 for the six pointings of
$\sim$\,0$\farcs$66$\times $\,0$\farcs$47 (PA\,$\sim$\,78\,deg.).  The
resulting continuum maps were created with the \textsc{clean}
algorithm using multi-frequency synthesis mode with a natural
weighting to maximise sensitivity. We initially created a dirty image
from the combined SPWs for each field and calculated the rms noise
values. The fields are then initially cleaned to 3$\sigma$ and then
masking boxes are placed on sources $>$\,4$\sigma$ and the sources
cleaned to 1.5$\sigma$. The six fields were then combined to create a
final image for source detection with an rms $\sigma_{\rm 1.25mm} =
$\,48\,$\mu$Jy\,beam$^{-1}$ at its deepest point, shown in
Fig.~\ref{fig:alma_fov}.

%
%

\begin{figure*}
	\centerline{\includegraphics[width=0.9\textwidth]{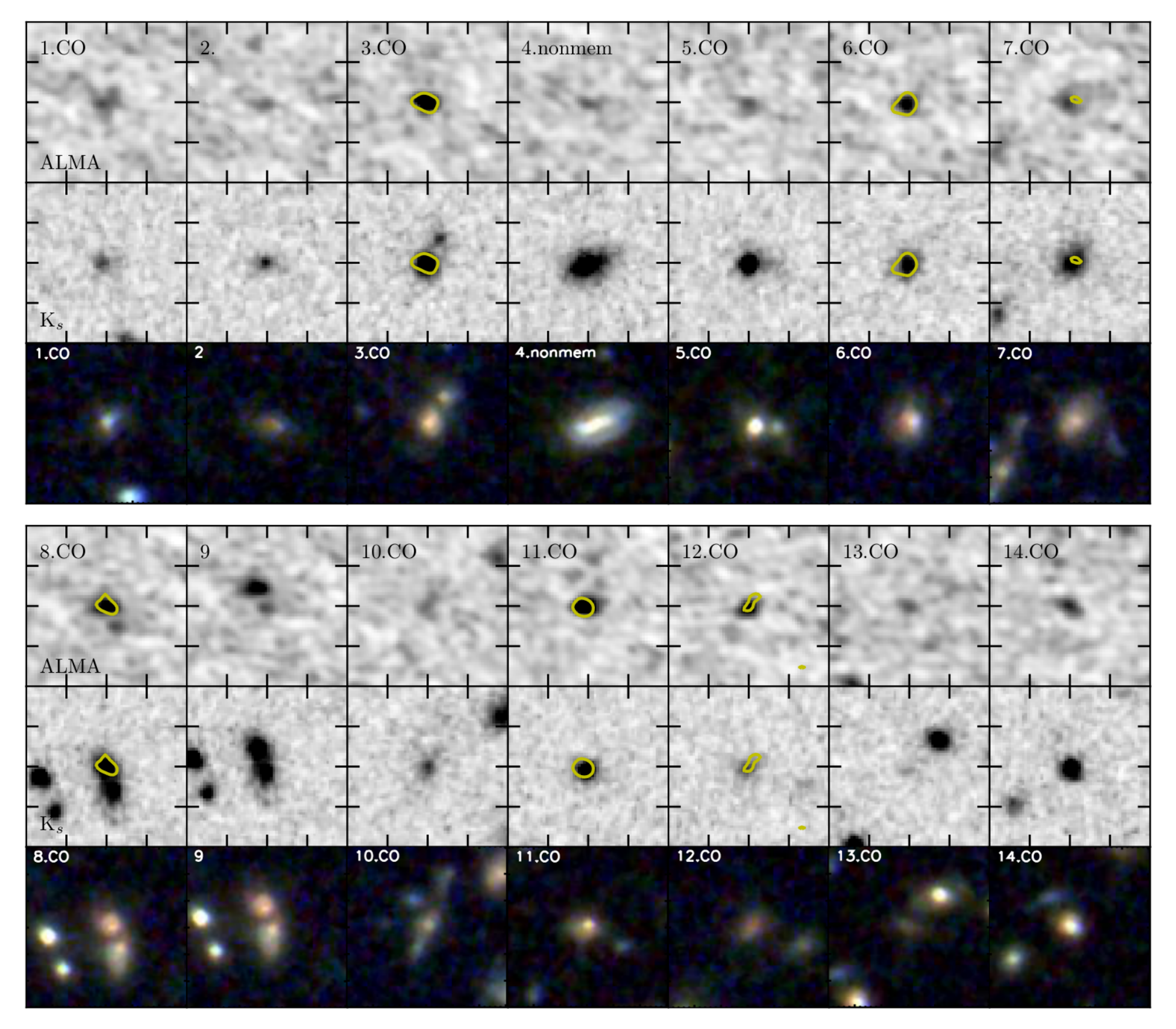}}
\caption{Thumbnails showing the ALMA Band 6 continuum (top row of each
  panel), $K_s$ (middle row of each panel) and three-colour {\it HST}
  WFC3 images (1.25, 1.40, and 1.60\,\micron, lower row of each panel),
  of the S/N\,$>$\,4 millimetre continuum sources detected in our
  map. Several sources display disturbed morphologies or very close
  neighbours (although these are typically faint in $K_s$ and hence
  likely to be low mass), suggesting that dynamical interactions may
  be triggering the strong star formation in these galaxies. Each of
  these thumbnails is centred on the positions given in
  Table~\ref{tab:results}. Six of the brightest sources in the ALMA
  continuum map: IDs 3, 6, 7, 8, 11, 12 additionally yield
  $^{12}$CO(2--1) and $^{12}$CO(5--4) emission-line detections, while
  IDs 1, 5, 10, 13, 15 have archival $^{12}$CO(2--1) detections from
  \cite{hayashi2017evolutionary} confirming that these are members of
  the cluster. Each thumbnail is 6\arcsec $\times$ 6\arcsec with major
  ticks every 1$\farcs$5 and with north up and east left. The yellow
  contours show the 3\,$\sigma$ contour for the integrated
  $^{12}$CO(5--4) map across the FWHM of the detected lines, showing
  the high-$J$ $^{12}$CO emission is aligned with the dust continuum.}
    \label{fig:b6dets}
\end{figure*}

\subsection{ALMA Band 3 Observations}

As well as the Band 6 mosaic, we also obtained a single pointing in
Band 3 centered on the cluster to cover $^{12}$CO(2--1) emission from
gas-rich cluster members. These observations were carried out on 2015
August 7 using thirty-nine 12\,m antennae (project ID:2013.1.01213.S), using
J2258$-$2758 as the bandpass calibrator, Ceres as the flux
calibrations, and the phase calibrator was J2206$-$1835 with an
on-science target integration time of 37.5\,minutes. Two spectral windows
were used centred at observed frequencies 93.246\,GHz and 95.121\,GHz,
with spectral bandwidths of 1.875\,GHz and a resolution of 1.938\,MHz
for both SPWs (corresponding to 6.1--6.2\,km\,s$^{-1}$). At this
observing frequency, the FWHM of the primary beam is
$\sim$\,61\arcsec\, and therefore the central $\sim$\,500\,kpc of the
cluster (Fig.~\ref{fig:alma_fov}) could be covered in a single
pointing. The same reduction approach was taken for the Band 3
observations as used for the Band 6 data, to create channel maps with
a velocity resolution of 50\,km\,s$^{-1}$ and a noise level in each
channel of 0.3--0.8\,mJy\,beam$^{-1}$.

\subsection{Source Detection}

To search for sources in the 1.25\,mm continuum map we used
\textsc{Aegean} \citep{hancock2012compact} to identify $>$\,4$\sigma$
detections. As part of this source extraction, we constructed a noise
map for the mosaic by deriving standard deviations of the flux density
in a box around each pixel with a size comparable to the synthesised
beam. Bright pixels are rejected in each box using a 3$\sigma$
clipping to avoid real sources contaminating the noise
map. The outside edge of the ALMA mosaic was then trimmed to
  the half-width half-maximum radius of the primary beams and source
  extraction was performed within this region which had a maximum
  noise of $\sigma_{\rm{rms}}=$\,0.09\,mJy\,beam$^{-1}$. Based on
this noise map we detect 14 S/N\,$>$\,4.0$\sigma$ candidate sources
from the Band 6 continuum map shown in Fig.~\ref{fig:alma_fov} and
listed in Table~\ref{tab:results}.  All continuum sources have
corresponding $K_{s}$, $i_{850}$, $r_{775}$ and $H_{160}$ band
counterparts within 0$\farcs$5, (Fig.~\ref{fig:alma_fov} \&
\ref{fig:b6dets}) and to estimate the reliability of these detections
we perform the same detection routine in the negative source map which
yields zero detections at $>$\,4$\sigma$.

We compare this number of detections with the blank-field 1.2\,mm
number counts of from the Hubble Ultra Deep Field
\citep{aravena2016alma,dunlop2016deep} (see also \cite{oteo2016almacal}).
For sources brighter than a flux limit of $\sim$\,0.18\,mJy, we would
expect $\sim$\,2\,$\pm$\,1 sources in the area of our continuum map.
Therefore, we appear to be detecting a $\sim$\,7$\times$ overdensity
of millimetre sources in the central projected 500\,kpc of XCS\,J2215
seen in Fig.~\ref{fig:alma_fov}.

To search for $^{12}$CO emission lines we first adopted a targeted
search by extracting spectra from the Band 3 and 6 datacubes at the
positions of the fourteen 1.25\,mm continuum detections.  In addition, we
extract spectra at the positions of the 25 spectroscopic members from
H10, the 46 sources from H09 with photometric redshifts indicating
possible cluster membership and the 20 [O{\sc ii}] emitters from
\cite{hayashi2014mapping} that are within the footprint of the ALMA
observations.

\begin{figure*}
	\includegraphics[width=\textwidth]{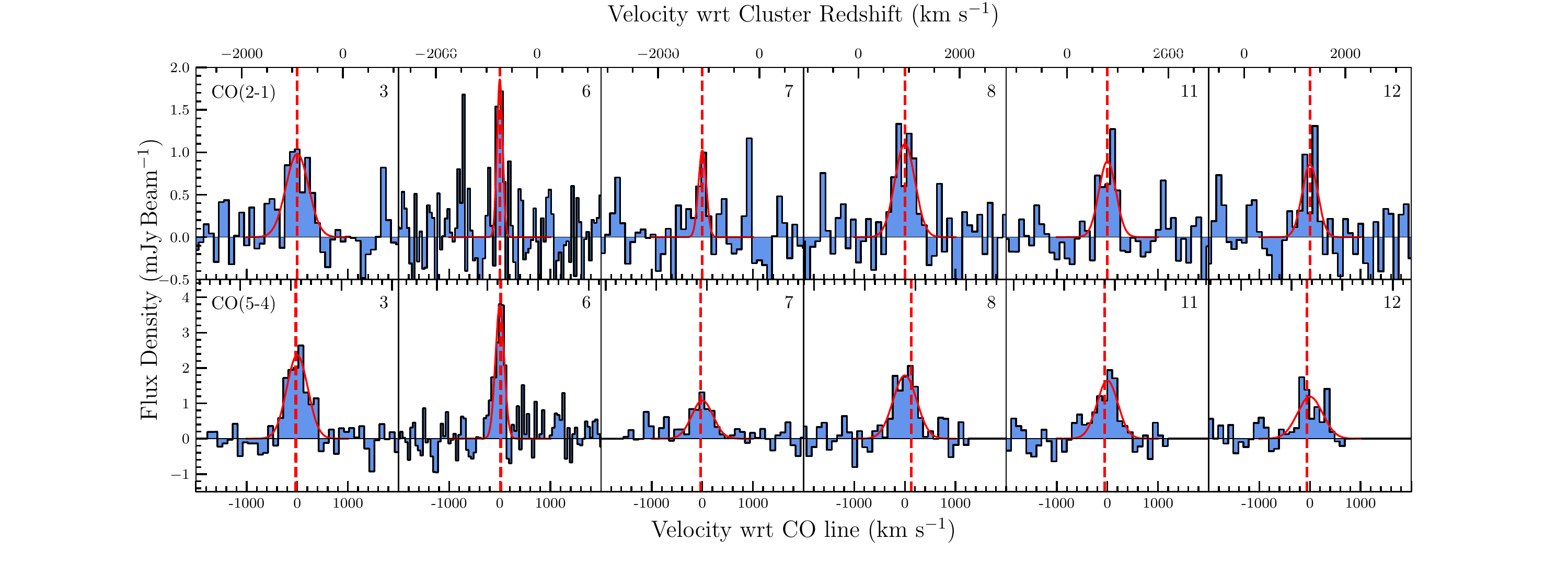}
    \caption{Spectra of the six $^{12}$CO detected cluster members
      with the top row showing the $^{12}$CO(2--1) emission in Band 3
      and the bottom row the $^{12}$CO(5--4) lines which fall in Band
      6.  In each case the best-fitting single Gaussian
        profiles are overlaid. We see detections of lines in both
      transitions in all six sources, confirming these millimetre
      sources as gas-rich cluster members. The lower velocity axis is
      centered on the peak velocity of the Gaussian fit to the
      $^{12}$CO(2--1) line whilst the upper velocity axis shows a
      velocity scale relative to the nominal cluster redshift of
      $z=$\,1.460. The spectra for IDs 3, 7, 8, 11, 12 are
        binned to 100\,km\,s$^{-1}$ resolution however for ID 6 the
        data was binned at 50\,km\,s$^{-1}$ due to their narrow line
        width. The lack of spectral coverage at velocities
      $>$\,1000\,km\,s$^{-1}$ for $^{12}$CO(5--4) IDs 8, 11 and 12 is
      due to the edge of the spectral window. The dashed red vertical
      lines show the velocity at which the $^{12}$CO(2--1) Gaussian's
      peak for each of our six detections.}
    \label{fig:codets}
\end{figure*}

We detect six significant emission lines in the Band 3 data, all
corresponding to bright dust continuum sources (IDs: 3, 6, 7, 8, 11,
12 in Fig.~\ref{fig:alma_fov}). One of these sources, ID 6, also has a
redshift from H10: $z=$\,1.454, consistent with our $^{12}$CO-derived
measurement, \cite{hayashi2017evolutionary} also report
  $^{12}$CO(2--1) detections for all six of these sources with
  redshifts consistent with our own.  We identify all of these lines
as $^{12}$CO(2--1) from cluster members and plot the spectra for these
in Fig.~\ref{fig:codets}. Applying the same procedure on the Band 6
cube yielded significant detections of $^{12}$CO(5--4) from just the
same six sources and these are also shown in Fig.~\ref{fig:codets} and
the line emission is contoured over the continuum in
Fig.~\ref{fig:b6dets} showing the high-$J$ gas is colocated with
the rest-frame 500\,$\mu$m dust emission.  For the $^{12}$CO(5--4)
emission lines, we subtracted the continuum emission in the $uv$ data
using the \textsc{uvconstsub} task in \textsc{casa} and, by averaging
data across channels, created continuum-subtracted channel maps at a
velocity resolution of 50\,km\,s$^{-1}$ with an rms of
0.3--0.4\,mJy\,beam$^{-1}$.  We find no individually detected
$^{12}$CO(2--1) or $^{12}$CO(5--4) emission from any of the other
sources in the spectroscopic or photometric samples.

%
%

For the galaxies where $^{12}$CO emission lines were detected, we
calculated the intensity-weighted redshift. This was calculated for
both the $^{12}$CO(5--4) and $^{12}$CO(2--1) lines, and for both lines
for all six sources the derived redshifts were in excellent agreement,
as can be seen in Fig.~\ref{fig:codets}. The redshift values reported
in Table~\ref{tab:results2} are the means of the redshifts derived
from the two transitions.  Line widths are derived from fitting
Gaussian profiles to the binned spectra using scipy.curve\_fit in {\sc
  Python} weighted by the uncertainty in the spectra.

We attempted a blind search for CO emission lines in the Band 3 and 6
cubes by collapsing them in 300\,km\,s$^{-1}$ wide bins (similar to
the observed FWHM of the targeted detections) and stepping the bins in
100\,km\,s$^{-1}$ increments across a velocity range $-$2$\sigma_{v} <
v < +$2$\sigma_{v}$ where $\sigma$ was the given velocity dispersion
of the cluster from H10. \textsc{Aegean} was used to detect peaks in
these collapsed channel maps looking for $>$4$\sigma$ detections. This
procedure recovered the six previously identified line emitters but
did not uncover any additional blind detections. We do see
  hints ($<3.5$\,$\sigma$) of further $^{12}$CO(5--4) detections at
  the locations and redshifts of two of the $^{12}$CO(2--1) detections
  from \cite{hayashi2017evolutionary} (ALMA.B3.04 and ALMA.B3.08)
  however these are all faint and none of them make our selection cut
  and so they are not considered for the rest of the paper.

\subsection{MUSE AO Observations}
Observations of the central 1\,$\times$\,1\,arcmin of XCS\,J2215
were taken with the MUSE IFU during the science verification of the
wide-field adaptive optics (WFM-AO) mode on 2017 August 15.  In this
mode, the GALACSI AO system is used increase the Strehl ratio of the
observations by employing four laser guide stars (and a deformable
secondary mirror) to correct for the turbulent atmospheric ground
layer.  An $R$\,=\,16.1 mag star located 95$''$ from the field center
was used to correct for the remaining atmospheric tip-tilt.
In total, we observed the cluster core for 3.5\,ks, which was split into 
four 860\,s exposures, each of which was spatially dithered by
$\sim$\,5$''$.  Between exposures, the IFU was also rotated through
90\,degrees to improve the flat-fielding along slices in the final
datacube.  We used the standard spectral range, which covers
4770--9300\,\AA\, and has a spectral resolution of
$R$\,=\,$\lambda$\,/\,$\Delta\lambda$\,=\,4000 at
$\lambda$\,=\,9200\AA\, (the wavelength of the [O{\sc ii}] at the
redshift of the cluster sample) -- sufficient to resolve the [O{\sc
    ii}]$\lambda\lambda$3726.2,3728.9 emission-line doublet.

To reduce the data, we use the MUSE {\sc esorex} pipeline which
extracts, wavelength-calibrates, flat-fields the spectra, and forms
each datacube.  Each observation was interspersed with a flat-field
to improve the slice-by-slice flat field (illumination) effects.  Sky
subtraction was performed on each sub-exposure by identifying and
subtracting the sky emission using blank areas of sky at each
wavelength slice, and the final mosaics were then constructed using an
average with a 3\,$\sigma$ clip to reject cosmic rays, using point
sources in each (wavelength collapsed) image to register the cubes.

For each source in our 1.2\,mm catalog, we extract a spectrum by
integrating the datacube within a 0$\farcs$5 aperture and search for the
[O{\sc ii}] emission at the cluster redshift.  For all the galaxies
with CO detections, we detect the [O{\sc ii}] emission line with
signal-to-noise ranging from 5 to 100.  We fit the [O{\sc ii}] emission
doublet and derive redshifts that agree with the CO (within their
1\,$\sigma$ error) in all cases.  We report the [O{\sc ii}] emission-line 
redshifts in Table~1.

\section{Analysis and Results} \label{sec:results}

Our ALMA survey of the central regions of XCS\,J2215 has resolved the
overdensity of four submillimetre sources in the core of the cluster
into 14 separate 1.25\,mm continuum sources
(Fig.\ \ref{fig:alma_fov}).  The four brightest 1.25\,mm continuum
components each correspond to one of the SCUBA-2 sources discovered by
Ma15: our ID 11 corresponds to SCUBA-2 source \#4 from Ma15, ID 3 is
\#11, ID 8 is \#6 and ID 6 is \#13. Our ALMA-detected millimetre
sources also confirm the primary IDs proposed by Ma15 for these
single-dish sources, but their analysis did not identify any of the
other 10 sources detected by ALMA in this region, which contribute to
the sub/millimetre flux seen by SCUBA-2.

As noted above, the five brightest millimetre continuum sources (plus
the seventh) also exhibit $^{12}$CO(2--1) and $^{12}$CO(5--4) emission
with redshifts that place them within the cluster.  In addition, we
match the remaining fainter continuum sources to the
\cite{hayashi2017evolutionary} $^{12}$CO(2--1) catalog, finding five
further matches: IDs 1, 5, 10, 13, 14. Matching to the H10 redshift
catalogue finds that a further continuum source, ID 4, has a spectroscopic
redshift of $z=$\,1.301 making it an interloper in the foreground of
the cluster and we do not consider it a cluster member in the
following analysis. In the remaining two continuum-selected
  ALMA sources (ID 2 and 9) we do not detect any emission lines in the
  MUSE spectra, which covers 4770--9300\,\AA\ (which corresponds to
  $z$\,=\,0.28--1.50 for [O{\sc ii}] emission).  However, we note that
  both of these two galaxies in the H09 photometric redshift catalogue
  have redshifts that are consistent with being possible cluster
  members, although the absence of lines in the MUSE spectra either
  suggests they are highly obscured, or they lie at higher redshift
  than $z$\,=\,1.50.  For the 11 spectroscopically confirmed
  millimetre-selected cluster members we derive a rest-frame velocity
  dispersion of $\sigma= $\,1040$\pm$100\,km\,s$^{-1}$. This is
marginally higher than the $\sigma=$\,720$\pm $110\,km\,s$^{-1}$
determined by H10 for the cluster members within the cluster core.
This difference is not statistically significant ($\sim$2\,$\sigma$),
but the sense of the difference is consistent with the expectation
that the millimetre-selected sources are likely to be relatively
recently accreted galaxies which have yet to fully virialise.

Our ALMA observations provide precise positions for the sub/millimetre
emission and so unambiguously identify the counterparts in the optical
and near-infrared wavebands, as shown in Fig.~\ref{fig:alma_fov} and
Fig.~\ref{fig:b6dets}.  Over half of these sources have companions on
scales of $\sim$\,2\arcsec--3\arcsec, although more than half of these are
faint or undetected in the $K_s$ band, suggesting they have relatively
modest stellar masses.  Nevertheless, this is some indication that
close tidal interactions or minor mergers may be the trigger for the
starburst activity seen in this population.

%
%
\begin{table*}
\begin{center}
	\caption{Properties of the ALMA 1.25\,mm continuum detections in XCS\,J2215}
	\label{tab:results}
	\begin{tabular}{ccccccccc}
		\hline
		ID & R.A.\ & Dec.\ & $S_{1.25\rm{mm}}$ & $L_{\rm FIR}$ & SFR & $z_{p}$ & $z_{s}^\ast$ & $z_{MUSE}$ \\
			& \multispan{2}{ ~(J2000) }  & (mJy) &  (10$^{11}$\,L$_{\odot}$) & ($M_{\sun}$\,yr$^{-1}$) & (H09) & \\
		\hline
		1  & 22\,15\,58.75 & $-$17\,37\,40.9 & 0.46$\pm$0.09 & 3.5$_{-2.2}^{+1.9}$ &  50$_{-30}^{+30}$ & 1.44$_{-1.10}^{+1.10}$ & {1.451$^{\dag}$}  &   1.451 (1.464)$^{\dag\dag}$  \\ 
		2  & 22\,15\,59.17 & $-$17\,37\,41.9 & 0.49$\pm$0.08 & 2.8$_{-1.4}^{+1.3}$ &  40$_{-20}^{+20}$ & 1.15$_{-0.20}^{+2.80}$ & ...              &   ...    \\ 
		3  & 22\,15\,58.54 & $-$17\,37\,47.6 & 0.93$\pm$0.05 & 8.8$_{-2.2}^{+4.4}$ & 130$_{-30}^{+60}$ & 1.99$_{-0.57}^{+0.39}$ & {\bf 1.453}      &   1.454  \\ 
		4  & 22\,15\,59.98 & $-$17\,37\,50.5 & 0.21$\pm$0.04 & 3.6$_{-1.2}^{+2.1}$ &  50$_{-20}^{+30}$ & 1.25$_{-0.36}^{+0.11}$ & {\it 1.301}      &   1.302  \\ 
		5  & 22\,16\,00.40 & $-$17\,37\,50.6 & 0.37$\pm$0.07 & 1.7$_{-0.8}^{+0.6}$ &  20$_{-10}^{+10}$ & 1.33$_{-0.19}^{+0.69}$ & {1.451$^{\dag}$}  &   1.452   \\ 
		6  & 22\,15\,57.23 & $-$17\,37\,53.3 & 0.68$\pm$0.08 & 9.4$_{-1.7}^{+2.7}$ & 140$_{-30}^{+40}$ & 1.30$_{-0.48}^{+0.92}$ & {\bf 1.454}      &   1.454  \\ 
		7  & 22\,15\,57.30 & $-$17\,37\,58.0 & 0.46$\pm$0.09 & 3.6$_{-2.0}^{+1.9}$ &  50$_{-30}^{+30}$ & 1.35$_{-0.18}^{+1.00}$ & {\bf 1.450}      &   1.453  \\ 
		8  & 22\,15\,59.71 & $-$17\,37\,59.0 & 0.88$\pm$0.08 & 3.3$_{-1.1}^{+0.9}$ &  50$_{-20}^{+10}$ & 1.32$_{-0.35}^{+1.10}$ & {\bf 1.466}      &   1.468  \\ 
		9  & 22\,15\,59.69 & $-$17\,37\,59.7 & 0.28$\pm$0.05 & 6.7$_{-1.7}^{+2.5}$ & 100$_{-20}^{+40}$ & 1.50$_{-0.22}^{+0.81}$ & ...              &   ...    \\ 
		10 & 22\,15\,57.48 & $-$17\,37\,59.9 & 0.18$\pm$0.04 & 1.8$_{-1.3}^{+0.8}$ &  30$_{-20}^{+10}$ & 1.97$_{-0.61}^{+0.37}$ & 1.450$^{\dag}$    &   1.451   \\ 
		11 & 22\,15\,58.15 & $-$17\,38\,14.5 & 0.98$\pm$0.06 & 5.5$_{-2.8}^{+1.6}$ &  80$_{-40}^{+20}$ & 1.73$_{-0.36}^{+0.52}$ & {\bf 1.467}      &   1.467  \\ 
		12 & 22\,15\,59.78 & $-$17\,38\,16.7 & 0.60$\pm$0.09 & 2.4$_{-1.0}^{+1.4}$ &  40$_{-10}^{+20}$ & 1.54$_{-0.47}^{+0.86}$ & {\bf 1.472}      &   1.472  \\ 
		13 & 22\,15\,58.09 & $-$17\,38\,19.4 & 0.30$\pm$0.07 & 3.9$_{-0.6}^{+1.2}$ &  60$_{-10}^{+20}$ & 1.34$_{-0.66}^{+1.70}$ & 1.467$^{\dag}$    &   1.469   \\ 
		14 & 22\,15\,58.23 & $-$17\,38\,22.3 & 0.56$\pm$0.08 & 3.6$_{-1.9}^{+2.0}$ &  50$_{-30}^{+30}$ & 1.46$_{-0.32}^{+0.99}$ & 1.457$^{\dag}$    &   1.457   \\ 
		\hline
	\end{tabular}

        {\small $^\ast$ Spectroscopic redshifts in {\bf bold} are from $^{12}$CO emission described in this paper, confirmed non-members are in {\it italics}.\\
          \small $^\dag$ $^{12}$CO spectroscopic redshifts from Hayashi et al.\ (2017).
          \small $^\dag\dag$ There are two galaxies separated by $<$0$\farcs$5 and $\sim$\,1500\,km\,s$^{-1}$, both of which are detetced in CO and [O{\sc ii}].}
\end{center}
\end{table*}

\subsection{SED Fitting}

We estimate the far-infrared luminosities for each of our continuum
sources by fitting their far-infrared and submillimetre photometry
using a library of galaxy template SEDs from
\cite{chary2001interpreting,draine2007dust,rieke2009determining}.  We
use our 1.25\,mm continuum fluxes along with fluxes from the lower
resolution single-dish observations from SCUBA-2 at 450 and
850\,$\mu$m \citep{ma2015dusty} and archival {\it Herschel} PACS data
at 100 and 160\,$\mu$m \citep[see][]{santos2013dust}. Due to
the low resolution for the single-dish observations, the fluxes for the
individual sources were estimated by deblending these maps using the
method detailed in \cite{swinbank2013alma} using the ALMA detections
as positional priors as described in Ma15. We calculate the infrared
luminosity from integrating the best-fitting SEDs for each galaxy in
the wavelength range 8--1000\,$\mu$m \,and from this derived the
far-infrared luminosity assuming the sources lie at the cluster
redshift (Table~\ref{tab:results}).  The far-infrared luminosities
show a $\sim$\,50\% dispersion at a fixed 1.25\,mm flux, but the
formal error bars are consistent with a single ratio, and hence there
is no strong evidence for a variation in SED shape within our small
sample.  In particular, we note that we obtained $^{12}$CO detections
for six of the brightest 1.25\,mm continuum sources, only three of
which fall in the top five brightest sources based on the far-infrared
luminosities.  This may indicate that the far-infrared luminosities
may be less reliable than adopting a single representative SED model
and fitting this just to the 1.25\,mm continuum flux.  We also caution
that if there are systematic differences in the dust SEDs of galaxies
in high-density environments, e.g., due to stripping of diffuse cold
gas and dust components \citep{rawle2012discovery}, then this will not
be captured by the templates in our library.

We next estimate the star-formation rate from the far-infrared
luminosities using the \cite{kennicutt1998star} relation and assuming
a Chabrier IMF. For the 14 ALMA continuum sources we derive
$L_{\rm{IR}}$ in the range of
(1.7--9.1)\,$\times$\,10$^{11}$\,$L_{\odot}$ and a median
(3.6$_{-1.2}^{+2.1}$)\,$\times$\,10$^{11}$\,$L_{\odot}$ which
corresponds to SFRs of $\sim $\,20--140\,$M_\odot$\,yr$^{-1}$
(Table~\ref{tab:results}). The derived luminosities of $L_{\rm{IR}}
=$\,10$^{11}$--10$^{12}$\,$L_{\odot}$ classify these cluster galaxies
as LIRGs with the brightest on the ULIRG boundary.

Integrating the ongoing star formation in the
  spectroscopically confirmed millimetre-selected cluster members we
  derive a total SFR in the central $\sim$\,500\,kpc of the cluster of
  $\gs$\,700\,$M_\odot$\,yr$^{-1}$.  Including the
  photometrically identified members (but excluding ID 4), this
  increases to $\gs$\,840\,$M_\odot$\,yr$^{-1}$. This is comparable
to the total SFR estimated by Ma15 within $R_{200}=0.8$\,Mpc
($\sim$\,100$''$), even though that region is much larger than the
extent of our current ALMA survey of the central $R\leq 0.25$\,Mpc
(Fig.~\ref{fig:alma_fov}).  Thus, our results reinforces the claims
that XCS\,J2215 demonstrates a very rapid increase in the SFR density
in the central regions of clusters out to $z\sim$\,1.5 and beyond.

\subsection{CO Line Properties}

To derive $^{12}$CO line properties we fit single Gaussians to each of
the Band 3 and 6 emission spectra, which appear to provide adequate
descriptions of the observed line profiles
(Fig.~\ref{fig:codets}). Estimates of the line widths were taken from
the FWHM of the Gaussian fits, and the flux density of the $^{12}$CO
lines were determined by integrating the $^{12}$CO spectrum,

\begin{equation}
   I_{\rm{CO}}=\int_{-2\sigma}^{+2\sigma}I(v)dv,
	\label{eq:fluxdensity}
\end{equation}

where $\sigma$ was taken from the Gaussian fits. Then, the $^{12}$CO luminosities were calculated using the relation given in \cite{solomon2005molecular}:

\begin{equation}
    L^{\prime}_{\rm{CO}}=3.25\times10^{7}S_{\rm{CO}}\Delta v\nu^{-2}_{\rm{obs}}D^{2}_{L}(1+z)^{-3},
	\label{eq:solomon}
\end{equation}

where $L^{\prime}_{\rm{CO}}$ is the line luminosity in
K\,km\,s$^{-1}$\,pc$^{2}$, $S_{\rm{CO}}\Delta v$ is the observed
velocity-integrated flux density in Jy\,km\,\,s$^{-1}$, $\nu$ is the
observed frequency of the emission line in GHz and $D_{L}$ is the
luminosity distance in Mpc. The FWHM and $^{12}$CO flux densities for
both transitions are given in Table~\ref{tab:results2}. For
  simplicity in comparing to the literature we adopted the same values
  for the constants $\alpha_{\rm CO}=1$, radius of galaxy $R_{kpc} =
  7$\,kpc and the $L_{\textrm{CO(2-1)}}$/L$_{\textrm{CO(2-1)}}$ ratio
  of $0.84\pm0.13$ from \cite{bothwell2013survey} when deriving
  $M_{\rm gas}$ and $M_{\rm dyn}$.  We then list the estimated gas
masses ($M_{\rm gas}$) for the galaxies based on their $^{12}$CO(2--1)
luminosities and adopting $\alpha_{\rm CO}=\,$1 (following
\citealt{bothwell2013survey}). We also list the dynamical masses
($M_{\rm dyn}$) for a disk-like dynamical model with a 7\,kpc radius
and the average inclination for a population of randomly orientated
disks (again following \citealt{bothwell2013survey}). We note
  that the derived values are highly dependant on the value of
  $\alpha_{\rm CO}$ and due to the SFRs being lower
  than the ULIRG sources in \cite{bothwell2013survey}, a more
  Milky Way-like $\alpha_{\rm CO}\sim 4.4$ which has been claimed to
  be appropriate for less active high-redshift galaxy populations
  might be more applicable \citep{tacconi2013phibss} however a Milky Way-like $\alpha_{\rm CO}$ results in gas masses for two out of our six detections
  being greater than our calculated dynamical masses. We note that our dynamical
  masses, whilst consistent with independent stellar mass estimations shown below,
  are based on adopting a mean inclination angle for populations of randomly oriented disks
  and an adopted value for the galaxy radius of $R_{kpc} = 7$\,kpc however the two galaxies 
  with gas fractions $>$1 adopting $\alpha_{\rm CO}\sim 4.4$ are ID 6 and ID 7 (see Fig.~\ref{fig:b6dets}) 
  and from their HST morphologies do not appear ``edge-on;'' therefore, the inclination angle assumption 
  will not create an underestimation of their dynamical masses. To reconcile the unphysical
  gas fractions would require dynamical masses estimated with galaxies of radii $>$35\,kpc which
  is again unphysical and an order of magnitude greater than previous size estimators of $^{12}$CO-emitting regions \citep{engel2010most}. This suggests, that for at least these two galaxies, that
  a Milky Way-like $\alpha_{\rm CO}\sim 4.4$ is an inappropriate conversion factor to use and a lower
  value more typical for local ULIRGs is more appropriate.

The median gas mass for the six galaxies is $M_{\rm
  gas}=1.6\pm0.2\times 10^{10}$\,$M_\odot$ (or $M_{\rm gas}$=4.3--10.5
$\times 10^{10}$\,$M_\odot$ for $\alpha_{\rm{CO}}=4.36$),
the median dynamical mass is $M_{\rm dyn}=0.9^{+0.3}_{-0.6}\times
10^{10}$\,$R_{kpc}$\,M$_\odot$ ($M_{\rm dyn}=6^{+2}_{-4}\times
10^{10}$\,$M_\odot$ for $R_{kpc} = 7$\,kpc) and the median gas
fraction is relatively low at $f_{\rm gas}=0.3\pm0.3$. We estimate stellar masses 
for our 1.2\,mm continuum-selected galaxies using their \emph{Spitzer} imaging.  
In particular, we exploit the archival IRAC imaging of XCS\,J2215 to measure IRAC 
3.6\,$\mu$m magnitudes for our sources, deriving a median magnitude of 21.1\,$^{+0.1}_{-0.2}$.  
At $z$\,=\,1.45, this corresponds to rest-frame $H$-band, which is sensitive 
to the underlying stellar mass of a galaxy.  We exploit the {\sc magphys}-derived 
stellar masses of comparably luminous submillimeter-selected galaxies in the Extended 
Chandra Deep Field South from \citet{da2015alma} and apply their median rest-frame 
$H$-band mass-to-light ratio to our sample. We derive a median stellar mass of 
$M_{*}=4^{+2}_{-2}\times 10^{10}$\,$M_\odot$ which suggests a median gas fraction 
for our six sources with CO detections of $f_{\rm gas}=0.3\pm0.4$ for $\alpha_{\rm{CO}}=1.0$, 
which is in excellent agreement with the gas fractions derived from the dynamical masses, 
suggesting our choice of $R_{kpc} = 7$\,kpc is appropriate.

Combining
these gas masses with the SFRs, we estimate a median
gas consumption timescale of 200$\pm$100\,Myrs which is comparable to
the crossing-time of the cluster core. However, as noted above,
  this is highly dependent on the choice of $\alpha_{\rm CO}$, scaling
  linearly, therefore if a more Milky Way-like $\alpha_{\rm CO}$ is
  appropriate, then the consumption timescale increases to
  $\sim$\,800\,Myrs, which is comparable to timescales expected for
  similar main-sequence galaxies at this redshift
  ($\sim$\,1.1$\pm$0.2\,Gyrs, \citealt{genzel2015combined}). We note
that the fainter $^{12}$CO detections for the remaining millimetre
sources with archival $^{12}$CO detections indicate gas masses of $\ls
1\times 10^{10}$\,M$_\odot$ and this may reduce the median gas
consumption timescale for the whole population (although these fainter
members also tend to have lower SFRs,
Table~\ref{tab:results}).

For the two continuum sources without spectroscopic redshifts,
  from their calculated L$_{\rm IR}$ and based on the scatter in the
  L$^{\prime}_{\rm CO}$-L$_{\rm IR}$ relation shown in
  Fig.~\ref{fig:bothplots} it is plausible that we might not detect
  $^{12}$CO(5-4) for ID 9.  For ID 2, on the other hand,  the combination of its faint
  continuum detection and location close to the edge of the ALMA
  primary beam (see Fig.~\ref{fig:alma_fov}) points to a non-detection
  possibly being a result of insufficient sensitivity.

%
%
\begin{table*}
	\centering
	\caption{Emission-line properties for $^{12}$CO(2--1) and  $^{12}$CO(5--4) detections in XCS\,J2215 member galaxies}
	\label{tab:results2}
	\begin{tabular}{ccccccc}
		\hline
		ID & $I_{\rm{CO(2-1)}}$ & FWHM$_{\rm{CO(2-1)}}$ & $I_{\rm{CO(5-4)}}$ & FWHM$_{\rm{CO(5-4)}}$ & M$_{\rm gas}$ & M$_{\rm dyn}$ \\
			& (J\,km\,s$^{-1}$) & (km\,s$^{-1}$) &(J\,km\,s$^{-1}$) & (km\,s$^{-1}$) & ($10^{10}$\,M$_\odot$) & ($10^{10}$\,M$_\odot$)  \\
		\hline
		3 & 0.5$\pm$0.1 & 530$\pm$90 & 1.2$\pm$0.1 & 490$\pm$40 & 2.4$\pm$0.7 & 11$\pm$4 \\ 
		6 & 0.25$\pm$0.08 & 130$\pm$30 & 0.80$\pm$0.09 & 200$\pm$20 &  1.0$\pm$0.2 & 0.6$\pm$0.3\\
		7 & 0.19$\pm$0.09 & 190$\pm$120 & 0.6$\pm$0.1 & 490$\pm$60 &  1.3$\pm$0.2 & 9$\pm$5\\
		8 & 0.6$\pm$0.1 & 460$\pm$90 & 1.0$\pm$0.1 & 550$\pm$50 & 2.2$\pm$0.6 & 9$\pm$4\\ 
		11 & 0.4$\pm$0.1 & 390$\pm$90 & 0.8$\pm$0.1 & 490$\pm$50 &  1.6$\pm$0.5 & 6$\pm$3\\ 
		12 & 0.3$\pm$0.1 & 370$\pm$90 & 0.7$\pm$0.2 & 580$\pm$100 & 1.5$\pm$0.7 & 5$\pm$3\\ 
		\hline
	\end{tabular}
\end{table*}

\begin{figure}
	\centerline{\includegraphics[width=\columnwidth]{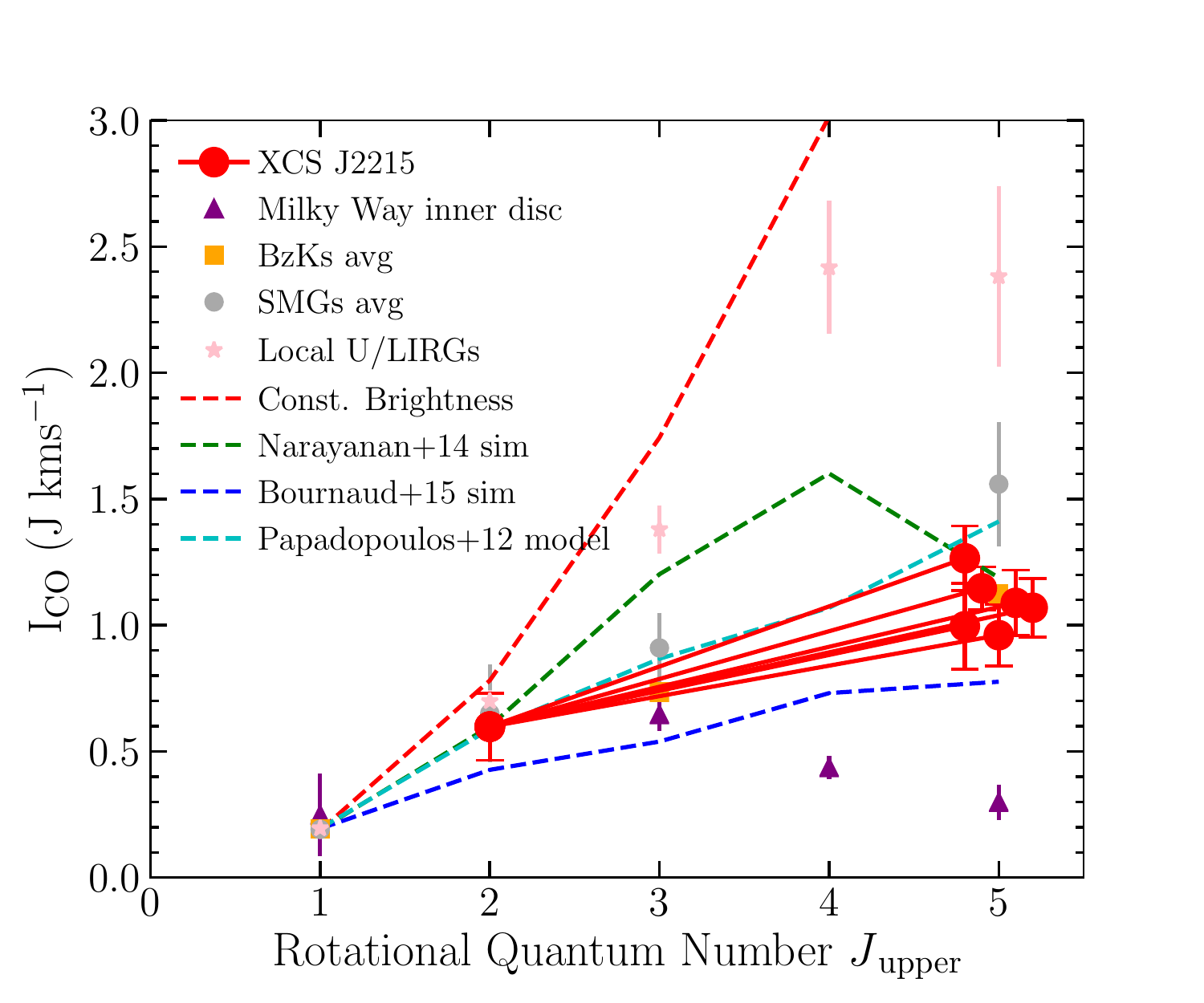}}
    \caption{$^{12}$CO SLEDs for the six XCS\,J2215 LIRGs compared to SLEDs for other populations from the compilation of  \cite{daddi2015co}. We see that our cluster LIRGs have SLEDs that peak at higher-$J$ than the Milky Way \citep{fixsen1999cobe},  indicating that the interstellar medium in these galaxies is more excited, although less excited than local U/LIRGs \citep{papadopoulos2012molecular}. Our sources appear to be similar to the submillimetre-selected field ULIRGs studied by  \cite{bothwell2013survey} and the less luminous BzK from \cite{daddi2015co}.  We also show model SLEDs from the simulations of \cite{narayanan2012general} and \cite{bournaud2015modeling}, and the toy model of \cite{papadopoulos2012molecular}. The latter implies that the interstellar medium is a two-phase mix of star-forming and non-star-forming gas, with 10\% of its gas in the star-forming phase.  All the SLEDs are normalised to the average $J=1$ transition for \cite{daddi2015co} $BzK$ average except for the Milky Way and XCS\,J2215 SLEDs which are normalised to the $BzK$ average $J=2$ transition. We note that if environmental processing has preferentially removed cool material from these galaxies, then their measured SLED will appear to be more ``active'' than it initially started off with.}
    \label{fig:sled}
\end{figure}

\begin{figure*}
	\plottwo{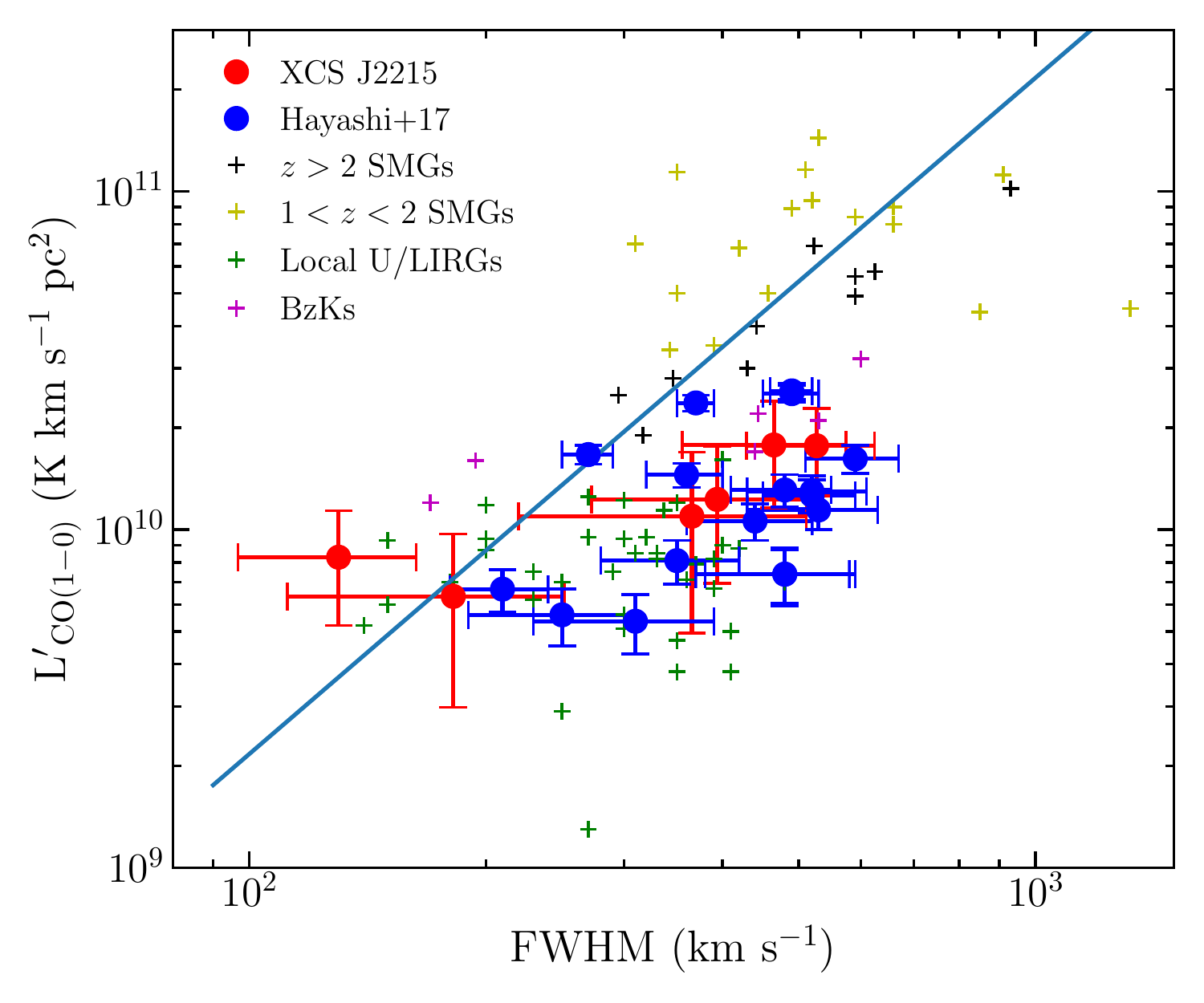}{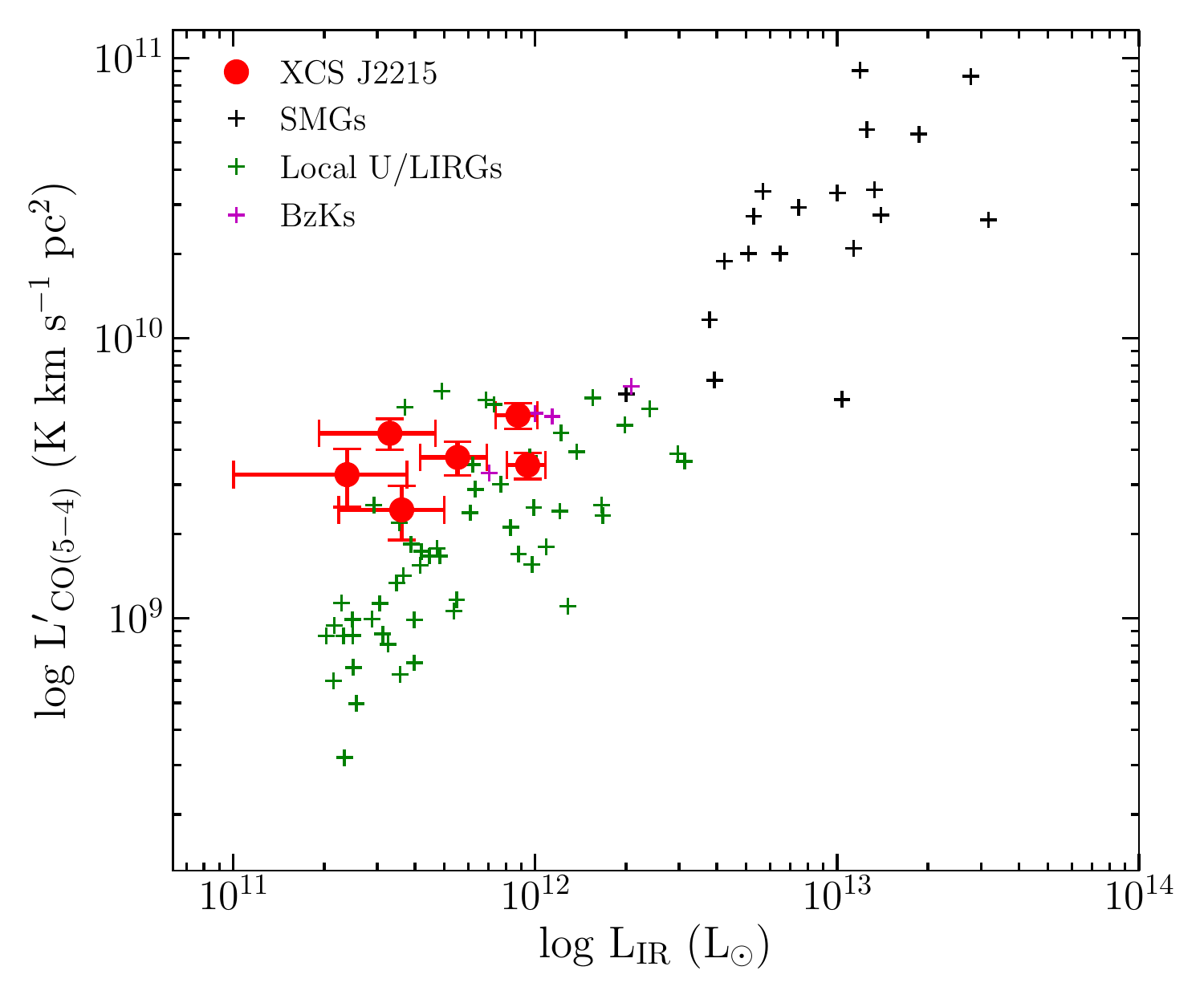}
    \caption{(a) Variation of $L^{\prime}_{\rm{CO(1-0)}}$  with FWHM of the line for the cluster LIRGs in this work (converting the $^{12}$CO(2--1) line luminosity and FWHM) compared to: the \cite{hayashi2017evolutionary} detections from the same cluster, local U/LIRGs \citep{downes1998rotating},  field SMGs from \cite{genzel2010study,bothwell2013survey} and $BzK$s from \cite{daddi2010very}. The solid line is the relation for $L^{\prime}_{\rm{CO(1-0)}}$ given in Eq.\,\ref{eq:lcofunc}. Our cluster LIRGs overlap with the local U/LIRG sample, although they appear to have slightly lower inferred $L^{\prime}_{\rm{CO(1-0)}}$ luminosities, at a fixed line width, compared  to submillimetre-selected field SMGs and BzKs at a similar redshift.  (b) The observed trend of $L^{\prime}_{\rm{CO(5-4)}}$ with $L_{\rm{IR}}$ for our six $^{12}$CO-detected cluster LIRGs and  comparison samples of local U/LIRGs, SMGs, and $BzK$s compiled by \cite{daddi2015co}. Our cluster LIRGs show $L_{\rm{CO(5-4)}}$/$L_{\rm{IR}}$ ratios consistent within the spread of the local U/LIRG population.}
    \label{fig:bothplots}
\end{figure*}

As we have observations of two $^{12}$CO transitions for our ALMA-identified cluster U/LIRGs, we can determine the ratio of the line brightness temperatures between the $^{12}$CO(5--4) and $^{12}$CO(2--1) transitions.  We show in Fig.~\ref{fig:sled} the  spectral line distributions (SLEDs) for our sources compared to other populations and models from the literature.   This shows that the cool interstellar medium within our cluster LIRGs is less excited than comparably luminous local galaxies, although it has very similar properties to that seen in   high-redshift, submillimetre-selected ULIRGs and $BzK$ galaxies.  To quantify this further, we determine a median value of the $^{12}$CO(5--4) and $^{12}$CO(2--1) line brightness ratio for our six sources of  $r_{54/21}=0.37\pm 0.06$.  We can compare this to the value derived for statistical samples of high-redshift, submillimetre-selected ULIRGs from \citep{bothwell2013survey} and $BzK$s from \cite{daddi2015co}:  $r_{54/10}=0.32\pm 0.05$, $r_{21/10}=0.84\pm 0.13$, which yield $r_{54/21}=0.38\pm 0.08$ for SMGs and $r_{54/10}=0.23\pm 0.04$, $r_{21/10}=0.76\pm 0.09$, which yield $r_{54/21}=0.32\pm 0.06$ for $BzK$s. As expected from Fig.~\ref{fig:sled}, these are in agreement to the values we derive and suggests comparable gas excitation in our sample of $z=$\,1.46 cluster LIRGs to the more luminous and typically higher-redshift field SMGs studied by \citet{bothwell2013survey}, as well as the less luminous $BzK$s.  This in turn suggests that the $r_{21/10}$ values for the  \citet{bothwell2013survey} sample should be broadly applicable to our sources.

%
%

%
%

\section{Discussion}

Our high-resolution continuum observations with ALMA have confirmed and significantly expanded the overdensity of luminous, dusty star-forming galaxies known in XCS\,J2215.  Our data also enable us to survey the cluster for massive gas reservoirs, and we find six gas-rich systems, associated with the typically brighter dust continuum sources.  These $^{12}$CO detections, along with five sources that have archival $^{12}$CO detections, unambiguously demonstrate that the majority of these galaxies are members of the cluster, while photometric redshifts suggest that two of the remaining continuum sources are also possible members.

We can use the CO line properties for our sources to compare to similar observations of other galaxy populations at high and low redshift to understand their physical properties.  Hence,  while $L^{\prime}_{\rm{CO}}$ provides a tracer for the molecular gas content in these galaxies, the FWHM of the emission lines provides us with a degenerate tracer of both the dynamical mass of the galaxy (narrower FWHM suggests lower mass) and inclination of the galaxy (narrower FWHM suggests  a more ``face-on'' galaxy). In Fig.~\ref{fig:bothplots}(a) we compare the $L^{\prime}_{\rm{CO(1-0)}}$ (converted from our $L^{\prime}_{\rm{CO(2-1)}}$ detections, adopting $r_{21/10}=$\,0.84) versus FWHM for our six $^{12}$CO-detected galaxies against a sample of local and high-redshift U/LIRGs. As in \cite{bothwell2013survey} we overlay the functional form given in Eq.~\ref{eq:lcofunc}

\begin{equation}
    L^{\prime}_{\rm{CO(1-0)}}=\frac{(V/2.35)^{2}R}{1.36\alpha G},
	\label{eq:lcofunc}
\end{equation}

where $V$ is the FWHM of the line, 1.36$\alpha$ is the $^{12}$CO to gas mass conversion factor, $R$ is the radius of the $^{12}$CO(2--1) emission region and $G$ is the gravitational constant. We, again, adopt the values of $\alpha=$\,1 and $R=$\,7\,kpc \citep{bothwell2013survey}. We see that LIRGs identified with ALMA in XCS\,J2215 fall within the scatter of the properties of local U/LIRGs, but slightly below the $BzK$s population seen at similar redshifts.  They may also show a marginally shallower trend than the functional form given in Eq.\,\ref{eq:lcofunc}, although the latter provides a good fit for the higher redshift and higher luminosity SMGs in \cite{bothwell2013survey}. We stress that the conversion of the line luminosities to $^{12}$CO(1--0) may result in systematic uncertainties between samples and individual sources in Fig.~\ref{fig:bothplots}.

Comparing the line widths for $^{12}$CO(2--1) and $^{12}$CO(5--4) emission lines for individual galaxies in our sample, we derive a median ratio of FWHM$_{21}$/FWHM$_{54}$\,=\,0.7\,$\pm$\,0.2.  If both transitions are tracing the rising part of the rotation curve in these galaxies, this marginal difference suggests that the $^{12}$CO(5--4) emission may be more extended than $^{12}$CO(2--1). This is the opposite behaviour to that expected if transitions with lower excitation temperatures have larger contributions from cool gas on the outskirts of galaxies \citep{,papadopoulos2001massive,ivison2011tracing,bolatto2013co}.  This could reflect environmental influences on the gas disks in these cluster LIRGs, with the removal of the more diffuse cool interstellar medium from their extended disks. Similar environmentally driven stripping of cooler material was invoked by \citet{rawle2012discovery} to explain the apparently higher dust temperatures seen in the SEDs of star-forming galaxies in  $z\sim$\,0.3 clusters.  We note that this would imply that before material was removed, the galaxies would have had a lower $r_{54/21}$ ratio than is currently observed, implying that originally they had a lower excitation SLED and a higher cold gas and dust mass and gas fraction.

As the far-infrared luminosity traces a galaxy's SFR and $L^{\prime}_{\rm CO}$  traces its gas content we show the ratio of these two observables for the ALMA-detected population in XCS\,J2215 in comparison to similar galaxies in the low- and high-redshift field in Fig.~\ref{fig:bothplots}b. To try to limit the effect of potential systematic errors we plot the line luminosities derived directly from our higher-S/N $^{12}$CO(5--4) detections, $L^{\prime}_{\rm{CO(5-4)}}$, and compare to similar high-$J$ observations of the other populations (following \cite{daddi2015co}).  Again, we see that our sources lie within the scatter of the local U/LIRG population, although they lie on the high side of the distribution. In comparison to the linear fit of the local U/LIRG population, our cluster galaxies show a median increase in $L^{\prime}_{\rm{CO(5-4)}}$ for their detected $L_{\rm{IR}}$ of 48\,\%$\pm$12\,\% (or conversely a deficit in $L_{\rm{IR}}$ at a fixed $L^{\prime}_{\rm{CO(5-4)}}$).  One possible explanation for this trend would be if the far-infrared luminosities of these sources are underestimated due to a relative paucity of cold dust (a consequence of the estimates of their far-infrared luminosities being driven primarily by the 1.25\,mm flux measurements due to their comparatively small errors in the SED fitting), due to environmental processing, compared to the template populations used to fit their SEDs (see \S 3.1).

\subsection{Environmental affects on the gas and dust in cluster U/LIRGs}

\begin{figure*}
	\plottwo{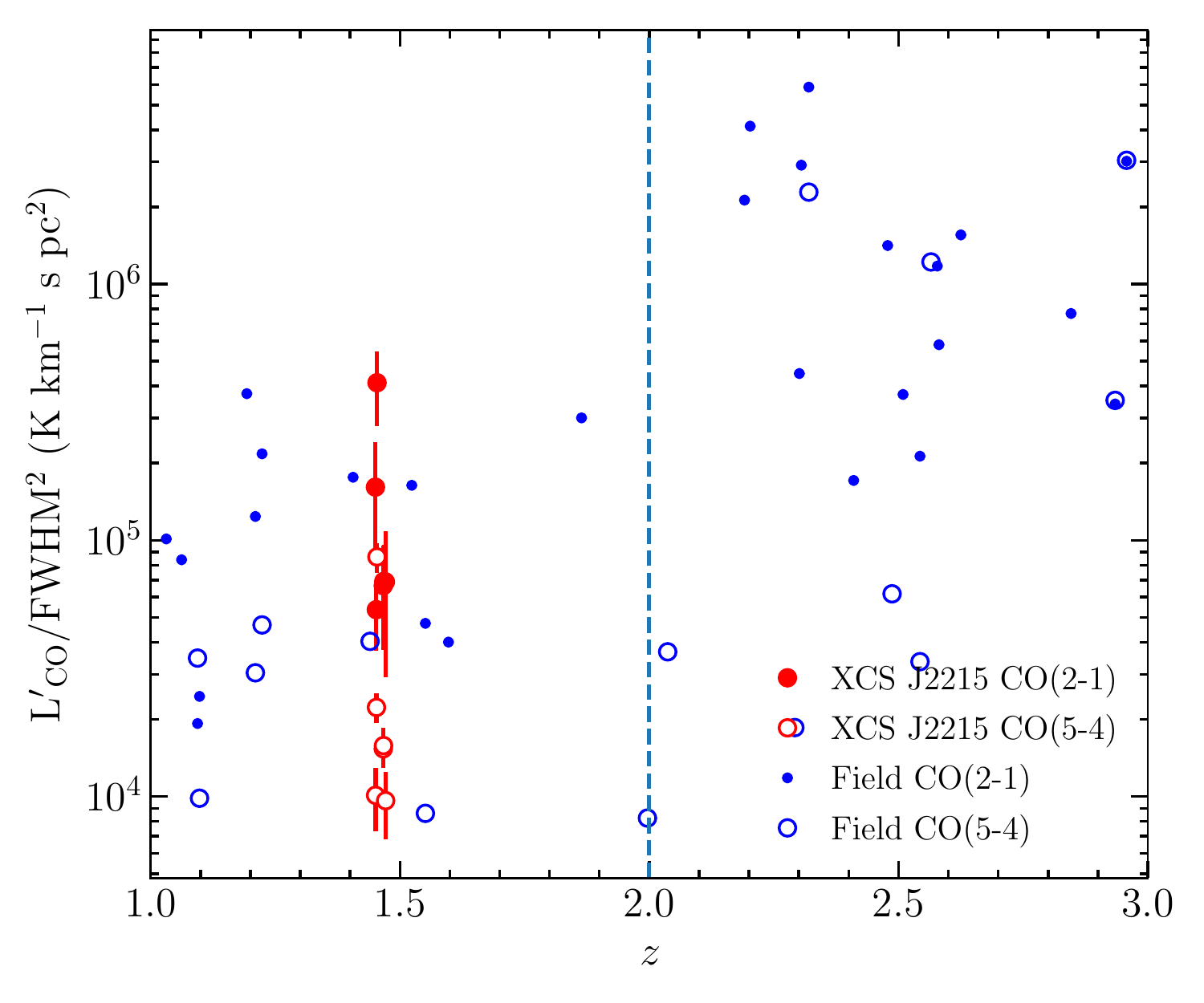}{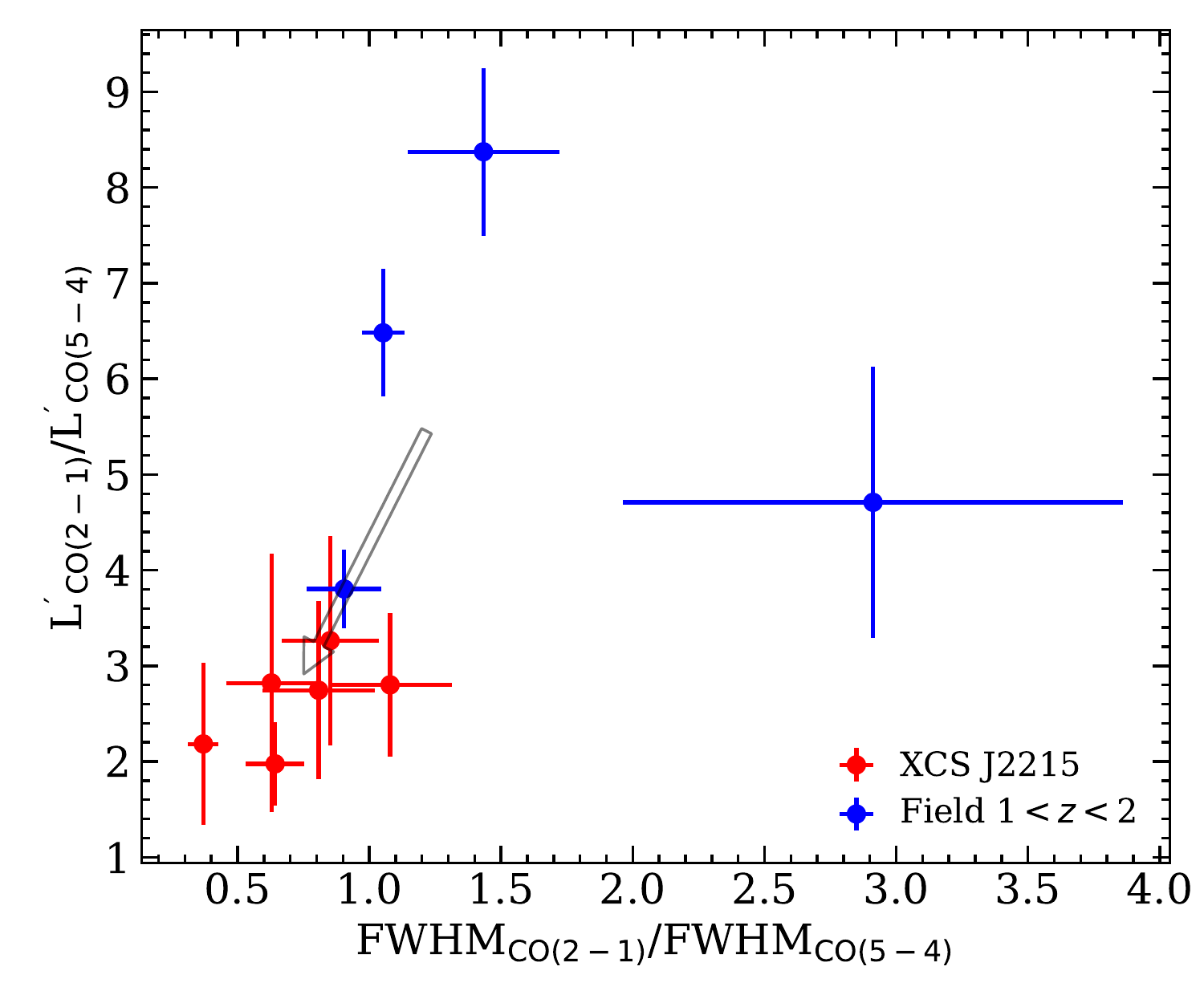}
    \caption{(a) Variation of $L^{\prime}_{\mathrm{CO}}$/FWHM$^{2}$ as a function of redshift. We take XCS\,J2215 cluster members and compare them to a field sample from the literature with published $^{12}$CO(2--1) and $^{12}$CO(5-4) observations. At the cluster's redshift, the galaxies with line detections show similar $L^{\prime}_{\mathrm{CO}}$/FWHM$^{2}$ ratios, within errors, to the field samples at their comparable redshift ($z\sim$\,1--2) for each CO transition. In the right panel, we show that it is not until you consider the ratio of $L^{\prime}$ and FWHM for the two transitions that the differences become clear. The dashed line represents our redshift cut for sample field galaxies used in the right-hand plot. (b) Plot of the ratios of the CO line luminosities against the corresponding FWHM of the two CO transitions. As in the left panel we plot the cluster members from XCS\,J2215 and comparison field sources at a similar redshift from the literature which have both low-$J$ CO detections ((2--1) or (1--0)) and high-$J$ CO detections ((7--6), (6--5), (5--4), or (4--3)) converted to $^{12}$CO(2--1) and $^{12}$CO(5--4), respectively, using the brightness ratios from \cite{bothwell2013survey}. We see that the cluster galaxies inhabit the lower left of the plot compared to the small sample of field sources with the relevant observations, suggesting that these galaxies may be comparatively poor in lower-density, cool gas in comparison to the field sample.   One possible explanation for this trend is environmental processes stripping of the cooler, less-bound gas from the outskirts of the galaxies. We overlay an arrow indicating the difference in the median values  of the field sample to the cluster galaxies to highlight the possible transition of a field galaxy to a cluster galaxy and the resulting effect on the low-$J$ CO line properties.}
    \label{fig:lowjstripping}
\end{figure*}

Looking at the gas and dust properties of our CO-detected galaxies in XCS\,J2215 we see several hints which all may be pointing to a relative paucity of cool gas and dust in these systems: (i) the galaxies typically have low $^{12}$CO(2--1) luminosities at a fixed FWHM,  compared to field galaxies; (ii) the line width measured from the $^{12}$CO(2--1) is typically smaller than that measured for $^{12}$CO(5--4); (iii) at a fixed  $^{12}$CO(5--4) line luminosity, these galaxies have lower inferred far-infrared luminosities (which is driven primarily by 1.25\,$\mu$m -- rest-frame $\sim$500\,$\mu$m -- flux) than comparable field galaxies. To isolate these trends in Fig.~\ref{fig:lowjstripping}a we plot $L^{\prime}_{\mathrm{CO}}$/FWHM$^{2}$, a proxy of gas fraction, as a function of redshift. The XCS\,2215 galaxies possess similar $L^{\prime}_{\mathrm{CO}}$/FWHM$^{2}$ for both the $^{12}$CO(2--1) and $^{12}$CO(5--4) transitions in comparison to similar U/LIRGs in the field taken from literature \citep{bothwell2013survey, carilli2013cool, zavala2015early, decarli2016alma}. While in Fig.~6(b), we consider the CO luminosity and FWHM {\it ratio} for these two transitions for the cluster galaxies and a sample of similar redshift field galaxies. We limit the redshift range for the field comparison sample to $z=1$--2 to try to remove evolutionary behaviour such as the increasing size of star-forming galaxies at lower redshifts \citep{van20143d} which might otherwise produce a spurious trend in L$^{\prime}_{\mathrm{CO}}$/FWHM$^{2}$ with redshift. Unfortunately, this leaves us with very few appropriate comparison sources as we are constrained by field galaxies in our redshift range that have observations in both high- and low-$J$ CO transitions. The CO luminosity and FWHM ratios both show a tentative trend that the cluster galaxies are comparatively poorer in the lower-density, cool $^{12}$CO(2--1) gas and also show a smaller $^{12}$CO(2--1) FWHM, suggesting that any deficit may be occurring on the outskirts of these galaxies. This would be consistent with the stripping of the cool, lower-density gas and dust from the gas disks as a result of an environment process (e.g.\ ram pressure stripping) that leaves the more tightly bound, denser $^{12}$CO(5--4) material relatively untouched \citep{rawle2012discovery}.  However, further observations of low- and high-$J$ CO are needed of larger samples of high-redshift cluster and field galaxies are needed to test this suggestion.

%
%

\subsection{Present descendants of cluster U/LIRGs}

The final issue we wish to address is, what are the likely properties of the present-day descendants of these galaxies?  They are bound in the cluster potential and so their stellar remnants will reside in a massive cluster of galaxies at the present day.  As we have noted, while these galaxies are rapidly forming stars at $z=$\,1.46 and $\sim$\,200--800\,Myrs later ($z\sim$\,1.2--1.4, depending upon the choice of $\alpha_{\rm CO}$)  this activity is likely to have declined substantially as their gas reservoirs are exhausted (this process will be even quicker if outflows or the further action of environmental processes suggested above accelerate to the removal of gas).  These star-formation events may form a significant fraction of the stellar mass of these systems, up to $\sim 10^{11}$\,$M_\odot$ (Ma15), although these estimates are highly uncertain.  Hence, we can conclude that the galaxies are likely to be massive at the present day and if their star formation terminates at $z\sim$\,1.2--1.4 then their stellar populations will appear old today as this corresponds to a lookback time of $\sim$\,9\,Gyrs.  

\begin{figure}
	\centerline{\includegraphics[width=\columnwidth]{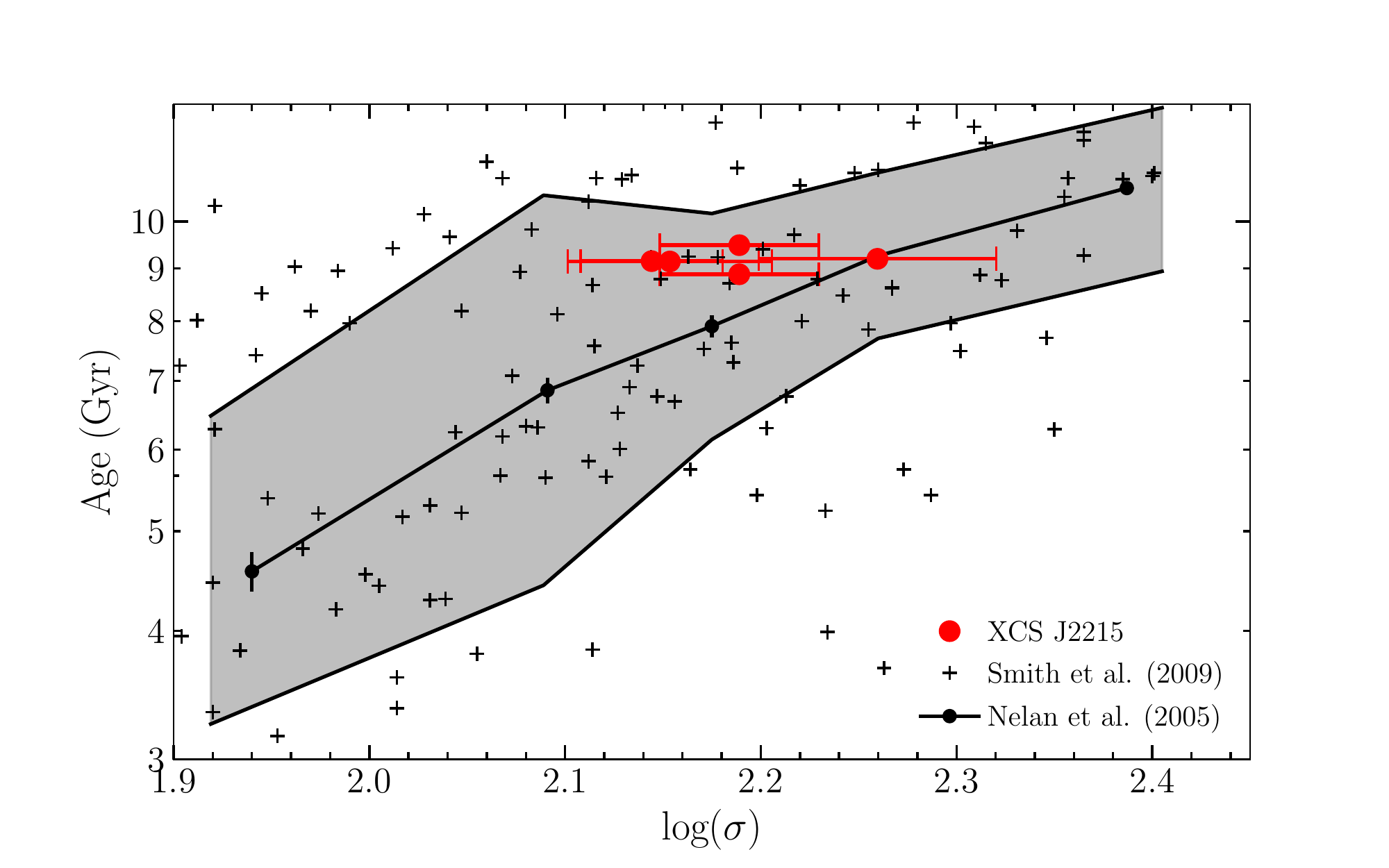}}
    \caption{A plot of the velocity dispersion of local early-type galaxies to their  luminosity-weighted stellar ages, adapted from
\cite{nelan2005noao}.  We show the median trend line and dispersion derived by \cite{nelan2005noao} and overplot measurements for individual
galaxies in the Shapley Supercluster from \cite{smith2009ages} to illustrate the scatter.  We  plot the velocity dispersions derived
from the Gaussian fits to the $^{12}$CO lines for the six CO-detected millimetre members in the core of XCS\,J2215, where their adopted
age is the lookback time to $z=$\,1.46, 9.3\,Gyrs.  These points therefore lie where they would appear today if the bulk of their stars
were formed in the starburst event we are currently witnessing.}
    \label{fig:nelan}
\end{figure}

Clusters of galaxies have long been known to house some of the oldest and most massive galaxies seen at the present day, but we can use our (relatively obscuration-free) measures of the dynamical masses of these galaxies to compare them more directly to local early-type galaxies. The lookback time to $z=$\,1.46 is 9.3\,Gyr, and our expectation is that the galaxies will rapidly exhaust their current gas supplies (and are unlikely to accrete substantial amounts of cold gas from their surroundings).  Hence, the stellar populations in their descendants at the present day are likely to have inferred ages of {\it at least} 9\,Gyr. In addition, if we assume the dynamical mass of the galaxies do not change significantly during the 9\,Gyr then the width of our Gaussian fits ($\sigma_{\rm Gauss}$) can be converted into an expected velocity dispersion ($\sigma$) by comparing the ratio of our mass estimator for disks with a simple virial equation estimator for a spherical mass distribution, giving us a conversion factor of $\sigma \sim 0.3\sigma_{\rm Gauss}$. We compare the expected properties of these galaxies to those of samples of early-type galaxies in local clusters in Fig.~\ref{fig:nelan}. We see that most (five out of six) of the $^{12}$CO-detected LIRGs in XCS\,J2215 have characteristics similar to those expected for the progenitors of relatively massive early-type galaxies at the present day. 

%
%

\section{Conclusions}

We have analysed ALMA 1.25\,mm and 3\,mm and MUSE-GALACSI observations
of a $\sim$\,500\,kpc diameter region in the core of the $z=$\,1.46
cluster, XCS\,J2215.  Our ALMA observations detect 14 luminous
1.25\,mm dust continuum sources within this region
(Fig.~\ref{fig:alma_fov}), representing a $\sim$\,7$\times$
over-density of sources compared to a blank field.  We detect line
emission from six of the brightest of these sources in the 1.25\,mm
and 3\,mm datacubes and associate these lines with redshifted
$^{12}$CO(5--4) and $^{12}$CO(2--1) transitions
(Fig.~\ref{fig:codets}).  These lines unambiguously identify the
millimetre sources as members of the clusters, while five other
continuum sources have archival $^{12}$CO(2--1) detections that also
place them in the cluster.  A further two sources have photometric
redshifts compatible with them being cluster members, but lack
spectroscopic redshifts from either ALMA or MUSE (consistent with the
expected field contamination in this map of $\sim$\,1--2 sources).
The 11 spectroscopically confirmed millimetre members have a
velocity dispersion which is marginally higher than the remainder of
the optical/near-infrared cluster members, hinting that they may be
relatively recently accreted onto the cluster.

Together, these results indicate that the vast majority of the millimetre sources are cluster members, and they confirm the intense obscured star formation occurring in the cluster core: $\gs $\,1000\,$M_\odot$\,yr$^{-1}$ in a $\sim$\,500\,kpc region, suggested by Ma15's earlier SCUBA-2 study. Combining our precise ALMA positions with high-resolution {\it HST} imaging, we see a high fraction of millimetre continuum-selected galaxies with close companions on $\ls$\,2\arcsec--3\arcsec scales (Fig.~\ref{fig:b6dets}), suggesting that galaxy--galaxy interactions may be a trigger for their activity, 
although most of these companions are faint in the $K_s$ band, indicating these are likely to be minor mergers/interactions.

We combine the $^{12}$CO(5--4) and $^{12}$CO(2--1) line fluxes for the cluster LIRGs to derive a median line brightness ratio,  $r_{54/21}=L^\prime_{\rm CO(5-4)}/L^\prime_{\rm CO(2-1)}=0.37\pm 0.06$.  This is comparable to the median ratio estimated for SMGs and $BzK$ populations at
similar and higher redshifts, indicating broadly similar gas excitation in our sample of $z=$\,1.46 cluster LIRGs to these
high-redshift star-forming populations (Fig.~\ref{fig:sled}).   We estimate gas masses (assuming $\alpha_{\rm CO}=$\,1) of $\sim$\,1--2.5\,$\times 10^{10}$\,$M_\odot$ and a median gas consumption time-scale of $\sim$\,200\,Myrs.  This time-scale is comparable to the time for a galaxy to cross the cluster core and so we anticipate that most of these galaxies will deplete their reservoirs before they exit the region they are currently seen in.

We also see a possible trend in terms of the gas and dust properties of the millimetre sources compared to  $z\sim$\,1--2 field galaxies, which may be pointing to a relative paucity of cool gas and dust in the cluster population:  (i) our XCS\,J2215 galaxies typically have lower $^{12}$CO(2--1) luminosities compared to their FWHM than comparable field galaxies,  (ii) the line width measured from the $^{12}$CO(2--1) is typically smaller than that measured for $^{12}$CO(5--4), (iii) at a fixed  $^{12}$CO(5--4) line luminosity, these galaxies have lower far-infrared luminosities than comparable field galaxies, and (iv) the ratio of the $^{12}$CO(2--1) and $^{12}$CO(5--4) CO line luminosities and the FWHM suggest that the cluster galaxies contain a larger fraction of warmer, denser $^{12}$CO(5--4) gas compared to field galaxies. These trends could be caused by the preferential removal of cooler, lower-density material as a result of an environmental process (possibly ram pressure stripping; \cite{rawle2012discovery}). Larger samples of cluster galaxies are needed to confirm the reality of this trend.

Finally, we have demonstrated that these galaxies have some of the properties of the expected progenitors of the massive, early-type galaxies which dominate the high-density regions of rich clusters of galaxies at the present day.  Specifically, their dynamical masses and stellar ages roughly match those seen in early-type galaxies in local clusters.

\acknowledgments

S.M.S.\ acknowledges the support of STFC studentship
(ST/N50404X/1). A.M.S.\ and I.R.S.\ acknowledge financial support from
an STFC (ST/L00075X/1). I.R.S.\ also acknowledges support from the ERC
Advanced Investigator program DUSTYGAL 321334, and a Royal
Society/Wolfson Merit Award. We thank Cheng-Jiun Ma, Chian-Chou Chen,
Roberto Decarli, Sune Toft, Tomotsugu Goto and Alasdair Thomson for
their help with the early stages of this project.  The ALMA data used
in this paper were obtained under program ADS/JAO.ALMA\#2015.1.00575.S
and ADS/JAO.ALMA\#2013.1.01213.S. ALMA is a partnership of ESO
(representing its member states), NSF (USA) and NINS (Japan), together
with NRC (Canada) and NSC and ASIAA (Taiwan), in cooperation with the
Republic of Chile. The Joint ALMA Observatory is operated by ESO,
AUI/NRAO, and NAOJ.  This paper used data from projects M13AU29 and
M13BU10 on the James Clerk Maxwell Telescope, which is operated by the
East Asian Observatory on behalf of The National Astronomical
Observatory of Japan, Academia Sinica Institute of Astronomy and
Astrophysics, the Korea Astronomy and Space Science Institute, the
National Astronomical Observatories of China and the Chinese Academy
of Sciences (Grant No.\ XDB09000000), with additional funding support
from the Science and Technology Facilities Council of the United
Kingdom and participating universities in the United Kingdom and
Canada.  The MUSE data are based on observations made with ESO
Telescopes at the La Silla Paranal Observatory under programme
60.A-9180.  All of the data used here is available through the ESO,
ALMA and \emph{HST} archives or in published papers.


\bibliographystyle{aasjournal}
\bibliography{ulirgbib}

\end{document}